\documentclass[structabstract]{aa}  
\usepackage{graphicx}
\usepackage{txfonts}
\usepackage{natbib}
\begin{document}
\title{{\it Suzaku} broad-band spectroscopy of RX~J1347.5--1145: constraints
  on the extremely hot gas and non-thermal emission}

% \subtitle{}

\author{N.~Ota \inst{1, 2}\fnmsep
\thanks{Research fellow of the Alexander von Humboldt Foundation},
        K. Murase \inst{3},
        T.~Kitayama \inst{4},
        E.~Komatsu \inst{5}, 
        M.~Hattori \inst{6},
        H.~Matsuo \inst{7},
        T.~Oshima \inst{8},
        Y.~Suto \inst{9},
        \and
        K.~Yoshikawa \inst{10} }

\institute{Max-Planck-Institut f\"{u}r  extraterrestrische Physik, 
              Giessenbachstra\ss e, 85748 Garching, Germany \\
                    \email{ota@mpe.mpg.de}
           \and
           Institute of Space and Astronautical Science, 
           Japan Aerospace Exploration Agency,
           3-1-1 Yoshinodai, Sagamihara, Kanagawa 229-8510, Japan
           \and
           Saitama University, Shimo-Okubo 255, Sakura, Saitama 338-8570, Japan
           \and
           Toho University, 2-2-1 Miyama, Funabashi, Chiba 274-8510, Japan
           \and
           The University of Texas at Austin, 2511 Speedway, RLM 15.306, Austin, TX 78712, USA
           \and
           Tohoku University, Aramaki, Aoba, Sendai 980-8578, Japan
           \and
           National Astronomical Observatory of Japan, 2-21-1 Osawa, Mitaka, Tokyo 181-8588, Japan
           \and
           Nobeyama Radio Observatory, Minamimaki, Minamisaku, Nagano 384-1805, Japan
           \and
           The University of Tokyo, Tokyo 113-0033, Japan
           \and
           Center for Computational Physics, University of Tsukuba, 
           Tsukuba, Ibaraki 305-8577, Japan 
           }

   \date{Received 5 May 2008; accepted 28 Aug 2008}

  \abstract
  {We present the results from the analysis of long Suzaku
    observations (149~ks and 122~ks for XIS and HXD, respectively) of
    the most X-ray luminous galaxy cluster, RX~J1347.5--1145, at
    $z=0.451$.}
  {In order to understand the gas physics of a violent cluster merger,
    we study physical properties of the hot ($\sim 20$~keV) gas clump
    in the south-east (SE) region discovered previously by the
    Sunyaev--Zel'dovich (SZ) effect observations. Using the hard X-ray
    data, a signature of non-thermal emission is also explored.}
  {We perform single as well as multi temperature fits to the Suzaku
    XIS spectra.  Then the Suzaku XIS and HXD, and the Chandra ACIS-I
    data are fit jointly to examine the properties of the hot gas
    component in the SE region. Finally, we look for non-thermal
    emission in the Suzaku HXD data.}
  {The single-temperature model fails to reproduce the 0.5--10~keV
    continuum emission and Fe-K lines measured by XIS
    simultaneously. The two-temperature model with a very hot
    component improves the fit, although the XIS data can only give a
    lower bound on the temperature of the hot component.  We detect
    the hard X-ray emission in the Suzaku HXD data above the
    background in the 12--40~keV band at the $9\sigma$ level; however,
    the significance becomes marginal when the systematic error in the
    background estimation is included.  With the joint analysis of the
    Suzaku and Chandra data, we determine the temperature of the hot
    gas in the SE region to be $25.3^{+6.1}_{-4.5}$ (statistical; 90\%
    CL) $^{+6.9}_{-9.5}$ (systematic; 90\% CL)~keV, which is in an
    excellent agreement with the previous joint analysis of the SZ
    effect in radio and the Chandra X-ray data. This is the first time
    that the X-ray analysis alone gives a good measurement of the
    temperature of the hot component in the SE region, which is made
    possible by Suzaku's unprecedented sensitivity over the wide X-ray
    band.  These results strongly indicate that RX~J1347.5--1145 has
    undergone a recent, violent merger.  The spectral analysis shows
    that the SE component is consistent with being thermal. We find
    the $3\sigma$ upper limit on the non-thermal flux,
    $F<8\times10^{-12}~{\rm erg\,s^{-1}cm^{-2}}$ in the 12--60~keV
    band, which provides a limit on the inverse Compton scattering of
    relativistic electrons off the CMB photons. Combining this limit
    with a recent discovery of the radio mini halo in this cluster at
    1.4~GHz, which measures the synchrotron radiation, we find a lower
    limit on the strength of the intracluster magnetic field
    $B>0.007~{\rm \mu G}$.}
   {}

\authorrunning{N. Ota et al.}
\titlerunning{{\it Suzaku} broad-band spectroscopy of RX~J1347.5--1145}

   \maketitle

\section{Introduction}\label{sec:intro}
\object{RX~J1347.5--1145} is one of the most extensively studied
distant clusters of galaxies, located at $z=0.451$, which is also
known as the most X-ray luminous cluster of galaxies on the sky. The
bolometric luminosity of \object{RX~J1347.5--1145} is $L_{\rm X,bol}
\sim 2\times10^{46}~{\rm erg\,s^{-1}}$ \citep{1997A&A...317..646S}.

The previous multi-wavelength observations have shown that
\object{RX~J1347.5--1145} has an unusually violent merger activity,
which makes this cluster an ideal target for probing the intracluster
gas physics and non-thermal phenomena associated with the cluster
merger at high redshifts.

The global temperature of the intracluster medium (ICM) of
\object{RX~J1347.5--1145} is as high as 9--14~keV, as indicated from
observations with several X-ray satellites \citep{1997A&A...317..646S,
  2001MNRAS.322..187E, 2002MNRAS.335..256A, 2007A&A...472..383G}.
This cluster has been classified as a ``cooling-flow'' cluster because
of its centrally peaked X-ray surface brightness profile as well as of
its high mass accretion rate estimated at the center
\citep{1997A&A...317..646S, 2002MNRAS.335..256A}.

The most striking feature of \object{RX~J1347.5--1145} is the presence
of an extremely hot, $\sim 20$~keV, gas clump in the south-east (SE)
region.  This component was discovered by radio observations of the
Sunyaev-Zel'dovich (SZ) effect towards \object{RX~J1347.5--1145} at
150~GHz as the prominent substructure of the SZ effect in the SE
region, about 10\arcsec off of the center \citep{2001PASJ...53...57K}.
\cite{2004PASJ...56...17K} have performed a detailed joint analysis of
the SZ effect decrement data at 150~GHz \citep{2001PASJ...53...57K},
increment data at 350~GHz \citep{1999ApJ...516L...1K}, and the {\it
  Chandra} ACIS-S3 X-ray data \citep{2002MNRAS.335..256A}, and
determined the temperature of the hot gas clump to be in excess of
20~keV, which is much higher than the average temperature of ambient
gas.

The coexistence of the cool and hot ICM in the central region
indicates a complex dynamical evolution of the system, such as a
recent merger.  In the optical band, the cluster center is dominated
by two cD galaxies. The dark matter distribution has also been studied
through the gravitational lensing effect
\citep[e.g.,][]{1997AJ....114...14F,1998ApJ...492L.125S,2002ApJ...573..524C,
  2002MNRAS.335..256A, 2008ApJ...681..187B,
  2008MNRAS.385..511M,2008A&A...481...65H}. An elongated, nearly
bimodal distribution was found \citep{2008ApJ...681..187B}, which also
points to the cluster merger.  Another line of evidence for the merger
comes from a recent discovery of the radio mini halo at the cluster
center \citep{2007A&A...470L..25G}, which suggests the presence of
relativistic particles powered by the merger as well as by the cooling
flow.

The X-ray data from the {\it ROSAT} satellite
\citep{1997A&A...317..646S} failed to identify the hot component in
the SE region, as its X-ray sensitivity did not extend to such a high
temperature. In fact, \object{RX~J1347.5--1145} had been considered as
a {\it relaxed}, non-merging cluster until the SZ effect revealed the
evidence for a violent merger. This example shows the importance of
having sensitivity to high temperatures well in excess of 10~keV.

Such a high temperature component seems common in merging clusters.
Another, perhaps most famous, example is the \object{Bullet cluster}
(\object{1E0657-56}), in which the hot gas substructure has also been
found \citep{2002ApJ...567L..27M}.  The analysis of the temperature
structure in merging clusters is a powerful tool for understanding the
gas physics in extreme conditions.  While the X-ray spectroscopy has
been used widely for this sort of study, it is not very easy to
precisely measure the temperature of very hot gas, as the majority of
X-ray imaging spectroscopic observations are limited to the energy
band below 10~keV, and thus they cannot measure a spectral cut-off at
$kT$ characteristic of the thermal bremsstrahlung emission.

To overcome this difficulty, \cite{2004PASJ...56...17K} used the fact
that the SZ effect is sensitive to arbitrary high temperature gas, and
performed a joint analysis of the SZ effect and X-ray images.  They
have determined the temperature of the SE excess component in
\object{RX~J1347.5--1145} to be $28\pm7$~keV, while the {\it Chandra}
X-ray spectroscopy, whose sensitivity degrades significantly beyond
7~keV, gave the lower limit of 22~keV.  This is one of the hottest gas
known to date. Based on the numerical simulations by
\cite{1999ApJ...520..514T}, \cite{2004PASJ...56...17K} suggested that
\object{RX~J1347.5--1145} has undergone a recent merger at the
collision speed of $4500~{\rm km\,s^{-1}}$, which is very similar to
what has been found more recently from the \object{Bullet cluster}
\citep{2007ApJ...661L.131M,2007MNRAS.380..911S,2007arXiv0711.0967M,
  2008MNRAS.384..343N}. This result is indicative of
\object{RX~J1347.5--1145} being essentially the \object{Bullet
  cluster}, except for the viewing angle.

The conclusion of \cite{2004PASJ...56...17K} relies on the SZ data
from high-resolution mapping observations of the SZ effect \citep[the
highest resolution ever achieved to
date;][]{1999ApJ...516L...1K,2001PASJ...53...57K}.  However, precision
of such SZ observations is still limited due to technological reasons;
thus, high-precision X-ray observations whose sensitivity goes well
beyond 10~keV are required to obtain reliable estimates of physical
properties such as the temperature and merging velocity.

The unprecedented sensitivity of the {\it Suzaku} satellite
\citep{2007PASJ...59S...1M} to hard X-ray emission well over 10~keV
offers a great opportunity for studying violent merger events.  In
this paper we present an analysis of the temperature structure and
non-thermal, high-energy component in the distant cluster
\object{RX~J1347.5--1145} with the on-board instruments, the X-ray
Imaging Spectrometer \citep[XIS;][]{2007PASJ...59S..23K}, and the Hard
X-ray Detector \citep[HXD;][]{2007PASJ...59S..35T,
  2007PASJ...59S..53K}. These instruments provide high-sensitivity
spectroscopic observations in a wide energy band up to several tens of
keV.  Combined with the {\it Chandra} data, we determine the
temperature of the SE clump {\it only from the X-ray spectroscopy},
and discuss its physical properties.  We also constrain, for the first
time, the non-thermal hard X-ray emission and estimate the magnetic
field strength in \object{RX~J1347.5--1145}.

Throughout this paper we adopt a cosmological model with the matter
density $\Omega_{M}=0.27$, the cosmological constant
$\Omega_{\Lambda}=0.73$, and the Hubble constant $H_0=72~{\rm
  km\,s^{-1}\,Mpc^{-1}}$, which are consistent with the WMAP 5-year
results \citep{2008arXiv0803.0586D,2008arXiv0803.0547K}.  At the
cluster redshift ($z=0.451$), $1\arcsec$ corresponds to
5.74~kpc. Unless otherwise specified, quoted errors indicate the 90\%
confidence range.

\section{Observation and data reduction}\label{sec:obs}
%%%%%%%%%%%%%%%%%%%%%%%%%%%%%%%%%%%%%%%%%%%%
\begin{table*}
\caption{Log of {\it Suzaku} observations of \object{RX~J1347.5--1145}. }\label{tab1}
\centering
\begin{tabular}{lllllll} \hline\hline
Target & Sequence No. & Date  &\multicolumn{2}{c}{Coordinates$^{\mathrm{a}}$} 
       & \multicolumn{2}{c}{Exposure [s]$^{\mathrm{b}}$} \\ 
 & & & RA [deg] & Dec [deg] & XIS & PIN \\ \hline
 RXJ1347.5-1145 & 801013010 & 2006-06-30 & 206.8560 & $-11.8093$ & 69661 & 56698 \\ 
 RXJ1347.5-1145 & 801013020 & 2006-07-15 & 206.8558 & $-11.8095$ & 79126 & 64922\\ \hline
\end{tabular}
\begin{list}{}{}
\item[$^{\mathrm{a}}$] Pointing coordinates in J2000.
\item[$^{\mathrm{b}}$] Net exposure time after data filtering.
\end{list}
\end{table*}
%%%%%%%%%%%%%%%%%%%%%%%%%%%%%%%%%%%%%%%%%%%%
%%%%%%%%%%%%%%%%%%%%%%%%%%%%%%%%%%%%%%%%%%%%
  \begin{figure}
   \centering
\rotatebox{0}{\scalebox{0.5}{\includegraphics{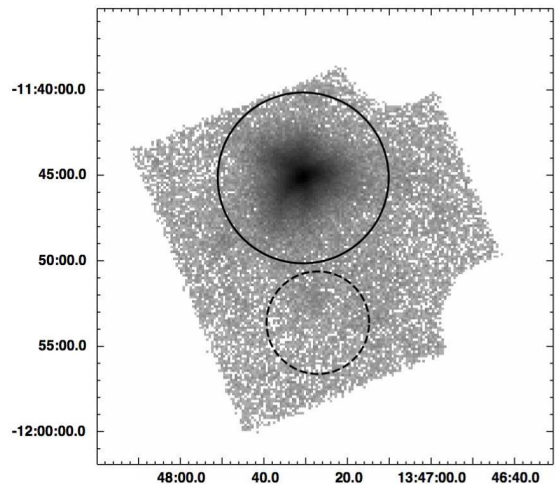}}}
\caption{{\it Suzaku} XIS-0 image of \object{RX~J1347.5--1145} in the
  0.5--10~keV band. The spectral integration regions for the cluster
  and the background are indicated with the solid and dashed circles,
  respectively. The two corners of the CCD chip illuminated by
  $^{55}$Fe calibration sources are excluded from the image. }
   \label{fig1}%
 \end{figure}
%%%%%%%%%%%%%%%%%%%%%%%%%%%%%%%%%%%%%%%%%%%%
Long observations of \object{RX~J1347.5--1145} were carried out twice
 in the {\it Suzaku} AO-1 period.  During the observations, XIS and
 HXD were operated in their normal modes.  To prioritize the HXD
 effective area, the target was observed at the HXD nominal aim-point
 (i.e., $3\arcmin.5$ off-axis relative to the XIS) . The total
 exposure time after data filtering is 149~ks/122~ks for XIS/HXD. The
 summary of the observations is given in Table~\ref{tab1}.

 The XIS consists of four X-ray CCD cameras: three front-illuminated
 CCDs (XIS-0, -2, -3: 0.4--12~keV) and one back-illuminated CCD
 (XIS-1: 0.2--12~keV), and covers a field of view of
 $18\arcmin\times18\arcmin$.  The ``Spaced Charge Injection'' option
 was not applied. The energy resolution was $\sim160$~eV at 6~keV
 (FWHM) at the time of the observations.  As shown in Fig.~\ref{fig1},
 the X-ray emission from the cluster is clearly detected. The X-ray
 peak position (13:47:30.3, $-11$:45:00.9) is consistent with that
 obtained from the {\it Chandra} (13:47:30.6,$-11$:45:09.3) within the
 accuracy of the attitude determination of {\it Suzaku}
 \citep[typically $<20\arcsec$; ][]{2008PASJ...60S..35U}.

 The non-imaging HXD covers a wide bandpass of 10--600~keV with PIN
 diodes (10--60~keV) and GSO scintillator (40--600~keV).  We shall use
 the PIN data in this paper.  The narrow field of view of $\sim
 30\arcmin\times 30\arcmin$ (FWHM) and the low background level of the
 PIN spectral data enable us to study the hard X-ray emission from the
 cluster up to several tens of keV, ideal for studying a violent
 merger event.  The bias voltage from one of four high-voltage units
 was reduced from the nominal value of 500~V to 400~V.  As a result
 the total effective area of the HXD/PIN decreased by about a few \%
 above 20~keV. This effect is incorporated in the detector response
 functions.

 We use the cleaned event files created by the pipeline processing
 version 2.0, and perform the data analysis using {\tt HEASOFT}
 version 6.3.1. The XIS data was filtered according to the following
 criteria: the Earth elevation angle $>10^{\circ}$, the day-Earth
 elevation angle $>20^{\circ}$, and the satellite outside the South
 Atlantic Anomaly (SAA). The HXD data was filtered according to the
 following criteria: the Earth elevation angle $>5^{\circ}$, the
 geomagnetic cut-off rigidity (COR) $>6$~GV, and the satellite outside
 the SAA.

 The XIS spectra were extracted from a circular region within a radius
 of $5\arcmin$ that is centered at the X-ray peak position (see
 Fig.~\ref{fig1}).  The background spectra were accumulated from the
 circular region within a $3\arcmin$ radius which is almost free from
 the cluster emission.  Since the background contributes 10--15\% to
 the total spectra, the systematic error due to the background
 uncertainty is small. To confirm this, we have checked the position
 dependence of the background spectra by analyzing the blank-sky data
 obtained during the $\sim 90$~ks North Ecliptic Pole observation on
 2006-02-10, and found that it is smaller than 7\% for all of the four
 sensors of XIS. Therefore, the background does not affect our
 spectral analysis within the statistical errors.

 The energy response files were generated by using the {\tt FTOOLS}
 task, {\tt xisrmfgen}. In order to take into account the vignetting
 effect of the X-ray telescopes \citep[XRT; ][]{2007PASJ...59S...9S}
 and a decrease of the low-energy efficiency due to the contaminating
 material on the optical blocking filter of the XIS, the auxiliary
 response files (arfs) were calculated by using {\tt xissimarfgen}
 \citep{2007PASJ...59S.113I}. Here, we assume that the X-ray surface
 brightness distribution of the cluster is represented by the double
 $\beta$-model: our reanalysis of the {\it Chandra} ACIS-I image
 yields the two core radii given by $r_{c1}=24$~kpc and
 $r_{c2}=94$~kpc, the same slope parameter, $\beta=0.63$, for both
 components, and a surface brightness ratio of $S_1/S_2 =8.1$ at the
 center (see Fig.~\ref{fig9}b).  Since the spatial distribution
   of \object{RX~J1347.5--1145} is compact compared to the typical
   size of the {\it Suzaku's} Point Spread Function (PSF), its
   detailed feature does not significantly affect our spectral
   analysis: if we compare the above arf files with those produced
   assuming a point-like source or an exposure-corrected {\it Chandra}
   image as an input source distribution, these three kinds of arfs
   agree within 3\% and give consistent spectral parameters within the
   present calibration uncertainty of the {\it Suzaku} telescope (see
   also Appendix~\ref{appendix:suzaku_chandra}).

 The hard X-ray spectra were accumulated from all units of the PIN
 diodes. The PIN detector background was subtracted by using the
 Non-X-ray Background (NXB) files\footnote{
   http://www.astro.isas.jaxa.jp/suzaku/analysis/hxd/pinnxb/pinnxb\_ver2.0/}
 provided by the HXD instrument team. Since precise modeling of the
 background is crucial for measuring the hard X-ray flux, we describe
 the method of background subtraction in \S\ref{sec:hxd_analysis}. In
 the spectral fitting, we use the PIN response
 function\footnote{ae\_hxd\_pinhxnome2\_20070914.rsp} that is
 appropriate for the epoch of observations.

\section{XIS analysis: 0.5--10~keV}\label{sec:xis_analysis}

In order to examine the global temperature structure of
\object{RX~J1347.5--1145}, we fit the APEC thermal emission model to
the observed XIS spectra in the 0.5--10~keV band.

The angular size of the X-ray emission is fairly compact and nearly
90\% of the total emission comes from the central $r<2'$ region, which
is comparable to the spatial resolution of the {\it Suzaku} XRT
\citep[Half Power Diameter $\sim2\arcmin$;][]{2007PASJ...59S...9S}.
We use the extraction radius of $5\arcmin$; thus, more than 97\% of
the photons from \object{RX~J1347.5--1145} are collected.

\subsection{Fitting with the single-temperature model}\label{subsec:xis_1apec}
%%%%%%%%%%%%%%%%%%%%%%%%%%%%%%%%%%%%%%%%%%%%
\begin{table*}
\centering
\caption{Single-temperature APEC model parameters fit to the XIS spectra
  taken by four 
  sensors (XIS-0, XIS-1, XIS-2, XIS-3), and the result from the
  simultaneous fit to all of the sensors (XIS-0, 1, 2, 3). } 
\label{tab2}
\begin{tabular}{lllll} \hline\hline 
 Sensor & $kT$~[keV] & $Z$~[solar] & $K^{\mathrm{a}}$ & $\chi^2/{\rm dof}$\\ \hline
XIS-0 & 13.31(12.76 -- 13.86) & 0.35(0.28 -- 0.40) & 1.37(1.37--1.39)$\times10^{-2}$ & 352.1/298 \\
XIS-1 & 12.23(11.54 -- 12.83) & 0.36(0.30 -- 0.41) & 1.41(1.39--1.44)$\times10^{-2}$ & 345.0/298 \\
XIS-2 & 12.87(12.40 -- 13.32) & 0.34(0.29 -- 0.38) & 1.40(1.38--1.42)$\times10^{-2}$ & 316.5/298 \\
XIS-3 & 12.76(12.28 -- 13.24) & 0.30(0.25 -- 0.35) & 1.45(1.43--1.47)$\times10^{-2}$ & 300.2/298 \\\hline
XIS-0,1,2,3 & 12.86(12.61 -- 12.94) & 0.33(0.31 -- 0.36) & 1.37(1.36--1.39)$\times10^{-2}$ & 1320.4/1198 \\\hline
\end{tabular}
\begin{list}{}{}
\item[$^{\mathrm{a}}$] The APEC normalization, $K=\int n_{\rm e}
  n_{\rm H} dV/(4\pi D_A^2(1+z)^2)~{\rm [10^{14}cm^{-5}]}$.
\end{list}
\end{table*}
%%%%%%%%%%%%%%%%%%%%%%%%%%%%%%%%%%%%%%%%%%%%
%%%%%%%%%%%%%%%%%%%%%%%%%%%%%%%%%%%%%%%%%%%%
\begin{table*}
\centering
\caption{The instrumental calibration parameters from the APEC model fit
 simultaneously to the XIS-0, 1, 2, and 3.}\label{tab3}
\begin{tabular}{llll} \hline\hline 
Sensor & Relative normalization & $N_{\rm c}~[10^{18}{\rm cm^{-2}}]$ & Gain offset [eV] \\ \hline
XIS-0  & 1(Fix)               & $0.012(< 0.116)$     & $-12.0$ ($-20.0$--$-6.0$) \\
XIS-1  & 1.035(1.023--1.047)  &  0.093(0.014--0.184) & $-16.0$ ($-22.5$--$-10.2$) \\
XIS-2  & 1.019(1.007--1.030)  &  0.011($<0.116$)     & $-9.8$ ($-16.5$--$-3.1$) \\
XIS-3  & 1.049(1.038--1.060)  &  0.000($<0.100$)     & $-20.1$ ($-26.4$--$-14.2$) \\\hline
\end{tabular}
\end{table*}%
%%%%%%%%%%%%%%%%%%%%%%%%%%%%%%%%%%%%%%%%%%%%
%%%%%%%%%%%%%%%%%%%%%%%%%%%%%%%%%%%%%%%%%%%%
\begin{figure*}
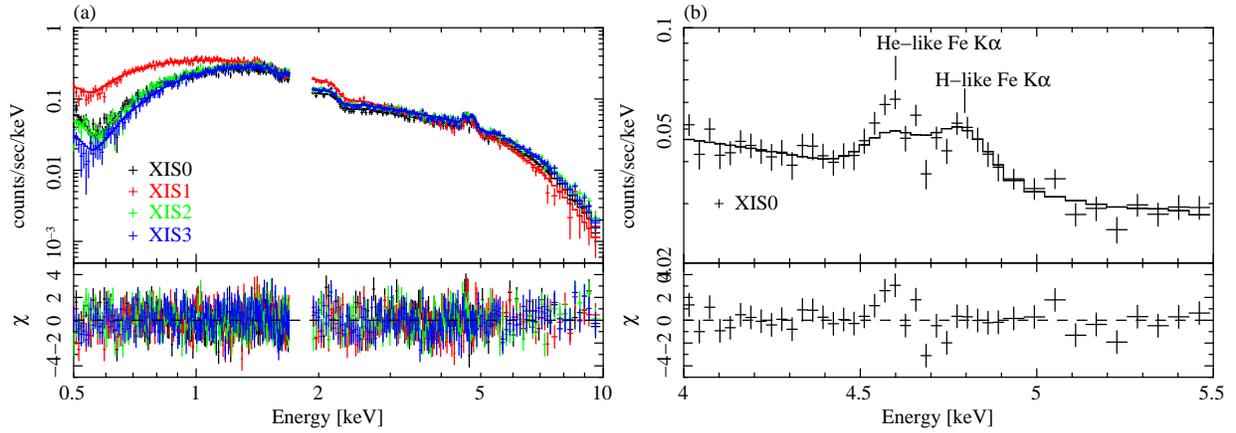

\centering
\rotatebox{270}{\scalebox{0.33}{\includegraphics{f2a.eps}}}
\rotatebox{270}{\scalebox{0.33}{\includegraphics{f2b.eps}}}
\caption{(a) Observed XIS spectra ($r<5\arcmin$). The spectra taken by
  four sensors, XIS-0 (black), 1 (red), 2 (green), and 3 (blue), are
  shown separately.  The solid lines in the upper panel show the
  best-fitting single-temperature APEC thermal plasma model
  simultaneously fit to all of the sensors, convolved with the
  telescope and the detector response functions.  In the bottom panel
  the residuals of the fit in units of the number of standard
  deviations are shown. (b) Blow-up of the XIS-0 spectrum in the
  4--5.5~keV band. The single-temperature model fails to fit the
  He-like Fe K$\alpha$ line.}
\label{fig2}
\end{figure*}
%%%%%%%%%%%%%%%%%%%%%%%%%%%%%%%%%%%%%%%%%%%%
%%%%%%%%%%%%%%%%%%%%%%%%%%%%%%%%%%%%%%%%%%%%
\begin{figure*}
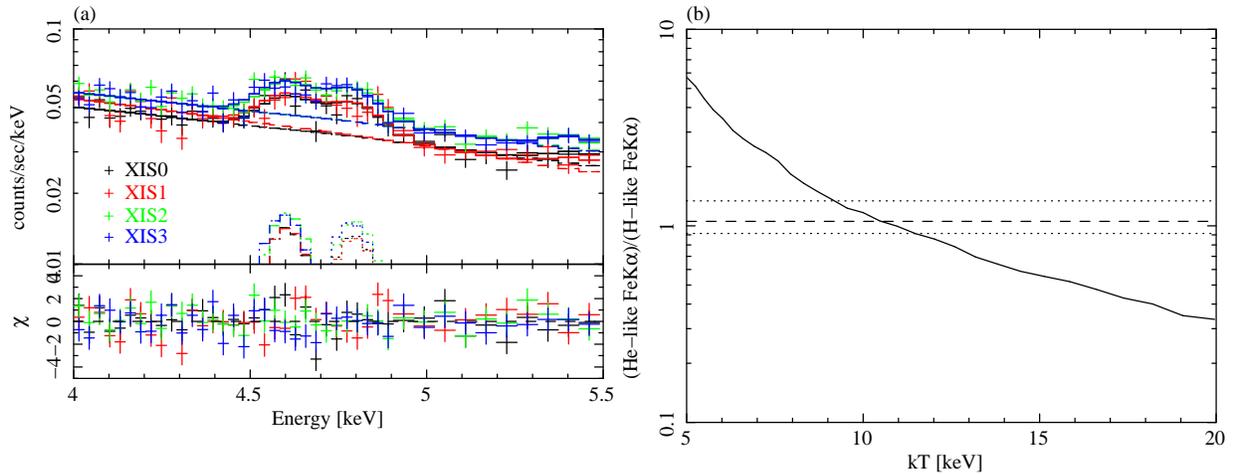

\centering
\rotatebox{270}{\scalebox{0.33}{\includegraphics{f3a.eps}}}
\rotatebox{270}{\scalebox{0.33}{\includegraphics{f3b.eps}}}
\caption{(a) Observed XIS spectra in the 4--5.5~keV band.  The spectra
  taken by four sensors, XIS-0 (black), 1 (red), 2 (green), and 3
  (blue), are shown separately.  The solid lines in the upper panels
  are the best-fitting continuum plus two Gaussian models.  In the
  bottom panel the residuals of the fit in units of the number of
  standard deviations are shown. (b) Line ratio (He-like Fe
  K$\alpha$)/(H-like Fe K$\alpha$) as a function of temperature,
  calculated from the APEC model. The horizontal dashed line and the
  dotted lines show the best-fitting line ratio and the 90\%
  confidence interval, respectively. }
\label{fig3}
\end{figure*}
%%%%%%%%%%%%%%%%%%%%%%%%%%%%%%%%%%%%%%%%%%%%
The XIS spectra with high photon statistics allow us to measure the
global temperature in two different ways: (1) the APEC model fit to
the 0.5--10~keV spectra, and (2) the use of the intensity ratio of two
Fe emission lines, namely (He-like Fe K$\alpha$)/(H-like Fe
K$\alpha$), as an indicator of the ``ionization'' temperature. By
comparing these two, we examine whether the emission can be modeled by
the single-temperature gas, or the data require more complex
temperature structure.

For the analysis (1) (APEC model), we fit the APEC model to the
observed 0.5--10~keV XIS spectra, where the redshift and the Galactic
hydrogen column density were fixed at $z=0.451$ and $N_{\rm
  H}=4.85\times10^{20}~{\rm cm^{-2}}$ \citep{1990ARA&A..28..215D},
respectively.  The metal abundances table in
\cite{1989GeCoA..53..197A} was assumed.  The spectral bins in the
1.7--1.9~keV range were excluded due to the large calibration errors
in the Si-edge structure.  In order to take into account the
uncertainty in the thickness of the contaminating material, the
absorption column densities of oxygen and carbon ($N_{\rm C}$ and
$N_{\rm O}$) were included in the spectral model with the ratio fixed
at $N_{\rm C}/N_{\rm O}=6$. In addition, there is a potential
uncertainty in the absolute energy scale on the order of 10--20~eV, as
indicated from the analysis of the $^{55}$Fe calibration source
spectra; thus, the gain offset for each XIS sensor is included in the
fitting model.

Since all of the four sensors yield consistent APEC parameters (see
Table~\ref{tab2}), we shall show the results from the simultaneous fit
to all of the sensors (XIS-0, XIS-1, XIS-2, and XIS-3) hereafter. When
performing the simultaneous fit, the normalizations of XIS-1, XIS-2,
and XIS-3 relative to that of XIS-0 were also treated as free
parameters. We find the temperature and the metal abundance of
$kT=12.86^{+0.08}_{-0.25}$~keV and $Z=0.33^{+0.03}_{-0.02}$~solar,
respectively.  The unabsorbed flux and luminosity in the 0.5--10~keV
band are $F_{\rm X,0.5-10~keV}=1.3\times10^{-11}~{\rm
  erg\,s^{-1}\,cm^{-2}}$, and $L_{\rm
  X,0.5-10~keV}=8.7\times10^{45}~{\rm erg\,s^{-1}}$, respectively. The
estimated bolometric X-ray luminosity is $L_{\rm
  X,bol}=1.37\times10^{46}~{\rm erg\,s^{-1}}$. In Table~\ref{tab3} we
list the the instrumental calibration parameters. In the following
analysis we shall use the best-fitting values for the relative
normalization factors, the carbon column densities, and the gain
offsets.

Fig.~\ref{fig2} shows the observed XIS spectra. Fig.~\ref{fig2}b is a
blow-up of the region around the redshifted He-like Fe K$\alpha$ and
H-like Fe K$\alpha$ lines at about 4.6~keV and 4.8~keV,
respectively. We find clear detections of both lines. Also shown is
the best-fitting single-temperature APEC thermal plasma model. The
single-temperature model fails to fit the He-like line. The reduced
$\chi^2$ from the single-temperature fit is $\chi^2=1320.4/1198$;
thus, the single-temperature model is rejected at the 99.3\% C.L.

We then attempt to fit the XIS spectra with the NEI (Non-equilibrium
ionization collisional plasma) model.  We found that the NEI model
does not improve the fit to the He-like Fe K$\alpha$ line and the
quality of the fit in the whole energy range is similar to the
12.9~keV APEC model: the resulting $\chi^2$ from the NEI model fit is
1340 for 1208 degrees of freedom. Moreover, the best-fit ionization
timescale of $>10^{12}~{\rm s\,cm^{-3}}$ is actually long enough for
the system to reach the equilibrium state \citep{1994ApJ...437..770M},
and thus the assumption that the system is in a non-equilibrium state
cannot be justified.  From these results we conclude that the
single-temperature NEI model fails to describe the observed XIS
spectra.

For the analysis (2) (line ratio), we model the observed XIS spectra
in the 4--5.5~keV band by the sum of the following components: (i) the
APEC model with the metal abundance reset to 0 for the continuum
emission, (ii) two Gaussian functions for the two major Fe lines at
4.6 and 4.8 keV, and (iii) another Gaussian line at 5.4~keV for the
blend of the He-like Ni-K$\alpha$ and He-like Fe-K$\beta$ lines.  In
Fig.~\ref{fig3}a we show the results of the fit.  This model gives an
acceptable fit to the data with $\chi^2/{\rm dof}=185/163$, or the
probability of finding higher $\chi^2$ values than observed is 11.4\%.

This model provides the line ratio, (He-like Fe K$\alpha$)/(H-like Fe
K$\alpha$), of $1.05(0.92-1.34)$. Fig.~\ref{fig3}b shows the
temperature dependence of the line ratio predicted by the APEC model,
folded with the XRT+XIS response functions. From this we find
$kT=10.4(9.1-11.4)$~keV, which is significantly different from that
found in the analysis (1), $kT=12.86(12.61-12.94)$~keV, suggesting
that the ICM cannot be explained by the single-temperature
model. Therefore, the ICM is more likely to be in the form of
multi-temperature plasma, which we shall explore in
\S~\ref{subsec:xis_2apec}.

Since the iron lines are emitted mostly from low-temperature gas (see
\S\ref{subsec:xis_2apec}), the line energies and widths are useful
probes of (i) the gas bulk motion and (ii) the turbulence in the
cluster core.  For (i), the observed centroid energies of the iron
lines are consistent with those expected from the $10$~keV APEC model
(6.69 and 6.97~keV in the rest frame) and the cluster redshift, to
within the 90\% statistical error of 10~eV.  As shown in
Table~\ref{tab3}, there is the gain offset of 15~eV when averaged over
the four XIS sensors, we simply assign the 90\% systematic error of
15~eV to the XIS energy scale. Then by adding the statistical and
systematic errors in quadrature, the 90\% upper limit on the
line-of-sight bulk velocity relative to $z=0.451$ is estimated as
$\Delta V< 1200~{\rm km\,s^{-1}}$. (ii) We find no significant line
broadening: the 90\% upper limit on the Gaussian width of the He-like
K$\alpha$ line is $\sigma_{\rm measured} <35$~eV. Since the measured
Gaussian width is the sum of the thermal Doppler broadening,
$\sigma_0$ ($\sigma_0=4$~eV for $kT=10$~keV), and the broadening due
to a turbulent velocity, $\sigma_{\rm turb}$, i.e., $\sigma^2_{\rm
  measured} = \sigma_0^2 + \sigma_{\rm turb}^2$, an upper limit on the
turbulent velocity dispersion is obtained as $\sigma_{\rm
  turb}<1500~{\rm km\,s^{-1}}$.  Therefore we did not find from the
analysis of the iron lines any significant bulk velocity and the
turbulent velocity, which is consistent with the view that the
low-temperature gas in the cluster core is relatively relaxed.  The
derived upper limits are also comparable to those obtained for the
core of the \object{Centaurus cluster} \citep{2007PASJ...59S.351O}.
If the velocity structure on a small ($<2\arcmin$) spatial scale
exists, the signals may be diluted due to the wide PSF of {\it
  Suzaku}. However, it is not easy to study the detailed spatial
structure in \object{RX~J1347.5--1145} because the angular size is
small.
  
\subsection{Fitting with the two-temperature model}\label{subsec:xis_2apec}
%%%%%%%%%%%%%%%%%%%%%%%%%%%%%%%%%%%%%%%%%%%%
\begin{table*}
\centering
\caption{Two-temperature APEC model parameters fit to the XIS
 spectra in the 0.5--10~keV and 4.0--5.0~keV bands.}\label{tab4}
\begin{tabular}{llllll} \hline\hline 
Energy band & $i$ & $kT_i$~[keV] & $Z_i$~[solar] & $K_i$ & $\chi^2/{\rm dof}$ \\ \hline
0.5--10~keV & 1 & 9.7(8.6--10.4)  & 0.33(0.31--0.36) & $1.07(0.88-1.13)\times10^{-2}$ & 1295.0/1207 \\
            & 2 & $>32.5$ & $=Z_1$  & $3.73(3.01-5.39)\times10^{-3}$ &             \\ \hline
4.0--5.5~keV& 1 & 9.8(3.6--11.6) & 0.35(0.28--0.52) & $9.2(2.9-14.1)\times10^{-3}$ & 185.2/159 \\
            & 2 & 34.4($>22.2$) & $=Z_1$ & $4.67(0.26-1.07)\times10^{-3}$ & \\ \hline
\end{tabular}
\end{table*}
%%%%%%%%%%%%%%%%%%%%%%%%%%%%%%%%%%%%%%%%%%%%
%%%%%%%%%%%%%%%%%%%%%%%%%%%%%%%%%%%%%%%%%%%%
\begin{figure*}
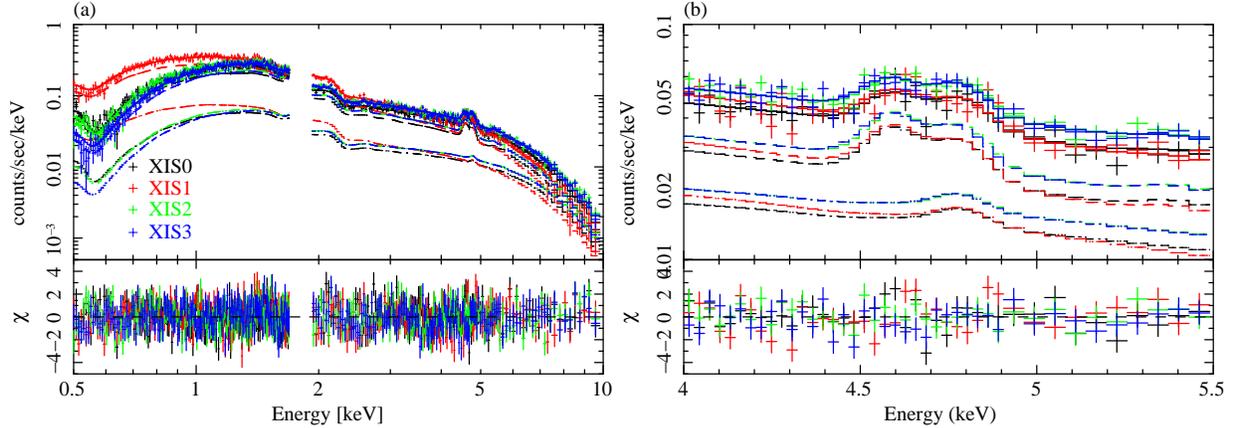

\centering
\rotatebox{270}{\scalebox{0.33}{\includegraphics{f4a.eps}}}
\rotatebox{270}{\scalebox{0.33}{\includegraphics{f4b.eps}}}
\caption{ (a) Two-temperature APEC model fit to the observed XIS
  spectra. The spectra taken by four sensors, XIS-0 (black), 1 (red),
  2 (green), and 3 (blue), are shown separately.  The solid lines in
  the upper panel show the sum of two APEC models with different
  temperatures, while the dashed lines show the low-temperature (upper
  curves) and high-temperature (lower curves) components separately,
  convolved with the telescope and the detector response functions.
  In the bottom panel the residuals of the fit in units of the number
  of standard deviations are shown. (b) Blow-up of the XIS spectra
  in the 4--5.5~keV band. The two-temperature model fits both the
  He-like and H-like Fe K$\alpha$ lines adequately.}
\label{fig4}
\end{figure*}
%%%%%%%%%%%%%%%%%%%%%%%%%%%%%%%%%%%%%%%%%%%%
Given that the single-temperature model fails to describe the observed
XIS spectra, we explore the two-temperature model consisting of two
APEC models with different temperatures, as the simplest case of the
multi-temperature plasma.  The results of the fit to the
two-temperature model are shown in Fig.~\ref{fig4} and
Table~\ref{tab4}, where the metal abundances of the two components are
assumed to be equal.

From the fit to the 0.5--10~keV spectra, we find that the
low-temperature component has the temperature of $kT_{\rm
  low}=9.7^{+0.7}_{-1.1}$~keV\footnote{The temperature of the
  low-temperature component is similar to the temperature determined
  from the spectra taken by the GIS detectors on board {\it ASCA} with
  the single-temperature model, $kT_{ASCA}=9.3^{+1.1}_{-1.0}$~keV
  \citep{1997A&A...317..646S}.}, whereas we only find a lower bound on
the temperature of the high-temperature component as $kT_{\rm
  high}>33$~keV.  This two-temperature model yields an acceptable
goodness-of-fit: the reduced $\chi^2$ is $\chi^2=1295.0/1207$, or the
probability of finding higher $\chi^2$ values than observed is 3.9\%.
In comparison to the single-temperature APEC model, the fit has
improved at more than the $99.99$\% confidence level according to the
F-test ($\Delta \chi^2=25$ for two additional parameters).
Restricting the energy range to 4.0--5.5 keV does not change the
parameters significantly (see the lower row of Table~\ref{tab4}).  The
reduced $\chi^2$ is $\chi^2=185.2/159$, or the probability of finding
higher $\chi^2$ values than observed is 7.6\%.

Since the observed Fe K$\alpha$ lines should be totally dominated by
the low-temperature gas (see Fig.~\ref{fig4}b), one
may compare $kT_{\rm low}=9.7^{+0.7}_{-1.1}$~keV and the temperature
derived from the line ratio, $kT=10.4(9.1-11.4)$~keV (see
\S~\ref{subsec:xis_1apec}), directly. These two estimates are in an
excellent agreement, which gives a support for the two-temperature
interpretation of the ICM of \object{RX~J1347.5--1145}.

Although the XIS spectra give the lower bound on the temperature of
the hotter component (see Table~\ref{tab4}), they fail to give the
upper bound. We find that this is due to the two-temperature model being
too simple to be realistic. More detailed modeling on the
hotter component, including a joint analysis with the {\it Suzaku} HXD data
and the spatially-resolved {\it Chandra} data, will be given in
\S~\ref{sec:xis+hxd_analysis}. 

\section{HXD/PIN analysis: 12--60~keV}\label{sec:hxd_analysis}
For the PIN diodes of HXD working in the energy band of 12--60~keV,
the background is composed of the following three components: (i) the
non-X-ray background (NXB), (ii) Cosmic X-ray Background (CXB), and
(iii) bright point-like sources within the detector field of views. Of
these, the NXB dominates the observed flux particularly for faint hard
X-ray sources such as clusters of galaxies.

Since the NXB has a significant time-variability, the reproducibility
of the NXB is the most important factor in the measurements of the
hard X-ray flux from clusters of galaxies. In \S~\ref{subsec:pin_nxb}
the systematic error due to NXB is carefully examined. In
\S~\ref{subsec:pin_cxb} the CXB model is calculated.  In
\S~\ref{subsec:pin_source} the point source fluxes are estimated from
the {\it XMM-Newton} observations of the same field.

\subsection{Non-X-ray background}\label{subsec:pin_nxb}
%%%%%%%%%%%%%%%%%%%%%%%%%%%%%%%%%%%%%%%%%%%%
\begin{figure*}
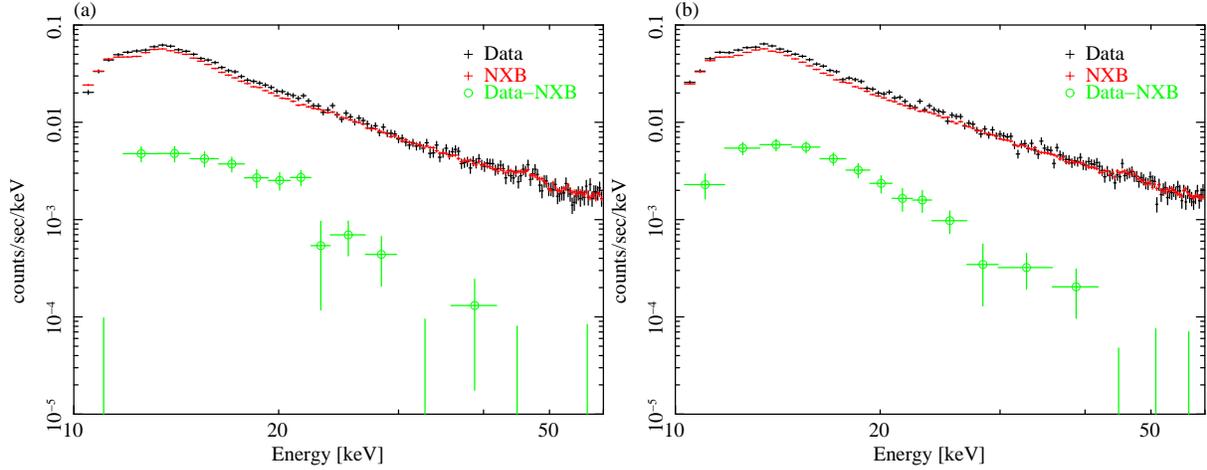

\centering
\rotatebox{270}{\scalebox{0.33}{\includegraphics{f5a.eps}}}
\rotatebox{270}{\scalebox{0.33}{\includegraphics{f5b.eps}}}
\caption{Observed PIN spectra during two observing periods of
  \object{RX~J1347.5--1145}. (a) Seq: 801013010, and (b) Seq: 801013020.  In
  each panel Data, NXB, and (Data$-$NXB) are shown with the black crosses,
  red crosses, and green circles, respectively.}
\label{fig5}
\end{figure*}
%%%%%%%%%%%%%%%%%%%%%%%%%%%%%%%%%%%%%%%%%%%%
%%%%%%%%%%%%%%%%%%%%%%%%%%%%%%%%%%%%%%%%%%%%
\begin{figure*}
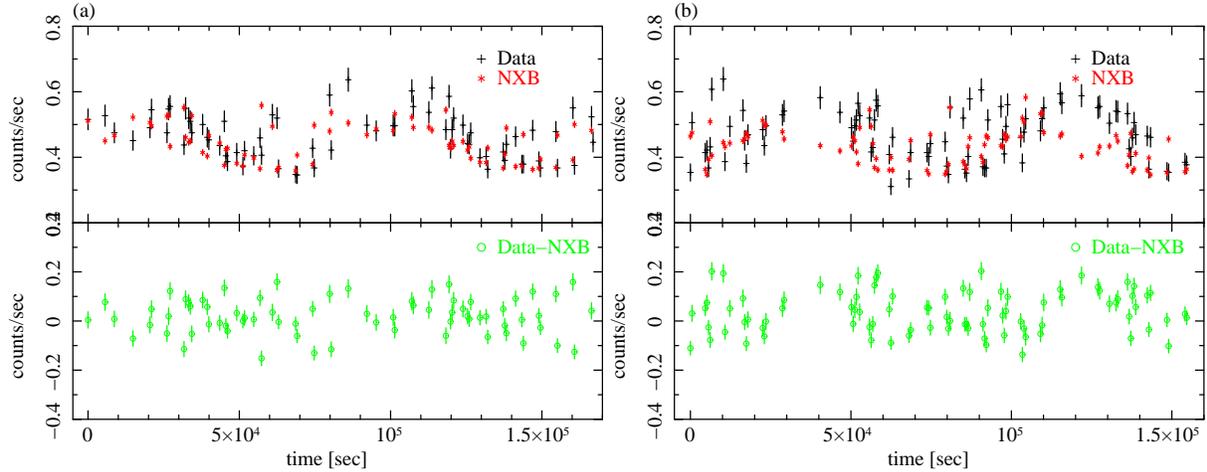

\centering
\rotatebox{270}{\scalebox{0.33}{\includegraphics{f6a.eps}}}
\rotatebox{270}{\scalebox{0.33}{\includegraphics{f6b.eps}}}
\caption{Observed light curves of the PIN 12--40~keV count rates
  during two observing periods. (a) Seq: 801013010, and (b) Seq:
  801013020. In the upper panels the Data and NXB rates are shown with
  the black and red crosses, respectively. In the bottom panels
  (Data$-$NXB) is shown with the green circles.  The error bars show
  the $1\sigma$ statistical errors.}
\label{fig6}
\end{figure*}
%%%%%%%%%%%%%%%%%%%%%%%%%%%%%%%%%%%%%%%%%%%%
%%%%%%%%%%%%%%%%%%%%%%%%%%%%%%%%%%%%%%%%%%%%
\begin{table*}
\centering
\caption{PIN count rates measured in the 12--40~keV and 40--60~keV bands.
 The quoted errors are $1\sigma$ statistical errors. }\label{tab5}
\begin{tabular}{lllllll} \hline\hline
Seq  &  Energy range & Exposure & Data  & NXB & (Data--NXB) & (Data--NXB)/NXB \\ 
         & & [s] &  [${\rm c\,s^{-1}}$] &  [${\rm c\,s^{-1}}$] &  [${\rm c\,s^{-1}}$] &  \% \\ \hline
801013010 & 12--40~keV &   56698 & $0.5180 \pm 0.0030$ & $0.4754 \pm 0.0009$ & $0.0426 \pm 0.0031$ & $8.96 \pm 0.67$  \\ 
801013020 & 12--40~keV &   64922 & $0.5223 \pm 0.0028$ & $0.4709 \pm 0.0008$ & $0.0514 \pm 0.0030$ & $10.92 \pm 0.63$  \\
801013010+801013020 & 12--40~keV &  121620 & $0.5203 \pm 0.0021$ & $0.4730 \pm 0.0006$ & $0.0473 \pm 0.0022$ & $10.00 \pm 0.46$  \\ \hline
801013010          &   40--60~keV &   56698 & $0.0497 \pm 0.0009$ & $0.0499 \pm 0.0003$ & $-0.0002 \pm 0.0010$ & $-0.44 \pm 1.96$  \\  
801013020          &   40--60~keV &   64922 & $0.0497 \pm 0.0009$ & $0.0495 \pm 0.0003$ & $0.0002 \pm 0.0009$ & $0.43 \pm 1.85$  \\    
801013010+801013020&  40--60~keV &  121620 & $0.0497 \pm 0.0006$ & $0.0497 \pm 0.0002$ & $0.0000 \pm 0.0007$ & $0.03 \pm 1.35$  \\    \hline  
\end{tabular}
\end{table*}
%%%%%%%%%%%%%%%%%%%%%%%%%%%%%%%%%%%%%%%%%%%%
The latest report from the instrument
team\footnote{JX-ISAS-SUZAKU-MEMO-2007-09 by Mizuno et al.} shows that
the reproducibility of the PIN NXB model for the data that have been
processed with the version 2 pipeline was typically 3\% ($1\sigma$).
This estimate is based upon systematic comparisons between the NXB
model and the PIN data during the Earth occultation in a trend
archive. We check below the validity of assigning the 3\% systematic
error to our PIN analysis.

In Fig.~\ref{fig5} we show comparisons between the observed PIN data
(hereafter ``Data'') and the NXB model in two observing periods, (a)
Seq: 801013010, and (b) Seq: 801013020. In Table~\ref{tab5} we
summarize the count rates in the 12--40~keV and 40--60~keV bands.  The
excess signal has been detected in Data$-$NXB at energies below 30~keV
during both observing periods.  The difference in Data$-$NXB between
two observing periods is well below the NXB itself: $({\rm Data}-{\rm
  NXB})_2-({\rm Data}-{\rm NXB})_1\sim 0.02 ({\rm NXB}_1+{\rm
  NXB}_2)/2$.  However, the difference is marginally ($\sim 2\sigma$)
inconsistent with zero compared to its statistical error: $({\rm
  Data}-{\rm NXB})_2-({\rm Data}-{\rm NXB})_1= 0.0088\pm 0.0043$,
which could be indicative of some residual time variability.

In general, the NXB intensity changes depending primarily on the
satellite's passage of the SAA and the COR
\citep{2007PASJ...59S..53K}. To study these effects, we bin the PIN
spectra under the following different conditions: elapsed time after
the passage of the SAA, TSAA$\gtrless6000$~s, and three different COR
[GV] ranges, $6<{\rm COR}<10$, $10<{\rm COR}<12$, $12<{\rm
  COR}<14$. This analysis shows that there is a trend to obtain a
higher count rate for TSAA$<6000$~s compared to TSAA$>6000$~s, and a
higher count rate for a smaller COR. However, the differences in the
count rates are below $\pm3$\% of the NXB intensity in each case.

In order to study the reproducibility of the NXB further, a comparison
was made for Data and NXB during the periods of the Earth occultation.
The Earth is known to be dark in hard X-rays and the Earth occultation
data can be regarded as NXB for the PIN observations.  As a result,
Data and NXB during the Earth occultation show a good agreement, and
Data$-$NXB is consistent with zero. From these studies we conclude
that the systematic error of the NXB model is fairly small
($\lesssim1$~\%).

In summary, the accuracy of the PIN NXB model is consistent with that
found in the latest report from the instrument team. Therefore, we
shall adopt the instrument team's estimate of the systematic error,
3\%, and propagate it through our spectral analysis of the HXD/PIN
data.

\subsection{Cosmic X-ray Background}\label{subsec:pin_cxb}
%%%%%%%%%%%%%%%%%%%%%%%%%%%%%%%%%%%%%%%%%%%%
\begin{figure}
\centering
\rotatebox{270}{\scalebox{0.33}{\includegraphics{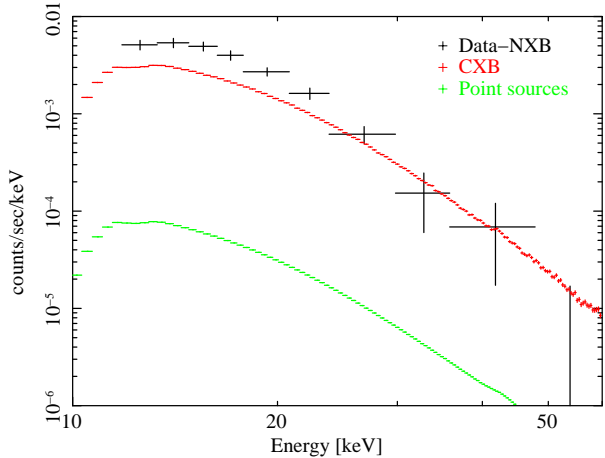}}}
\caption{Observed PIN spectrum minus the nominal NXB model
  (Data$-$NXB; black), the simulated CXB model (red), and the point
  source contribution estimated from the {\it XMM-Newton} observations
  of the same field (green).}
\label{fig7}
\end{figure}
%%%%%%%%%%%%%%%%%%%%%%%%%%%%%%%%%%%%%%%%%%%%
We calculate the CXB spectrum in the PIN band assuming a power-law
model with an exponential cut-off at 40~keV as previously reported
from the {\it HEAO-1} A2 \citep{1987IAUS..124..611B}.  Since
the PIN detector response assumes a uniform distribution of emission
over $2^{\circ}\times2^{\circ}$ on the sky, the CXB model for 4 square
degree field in units of ${\rm photons\,cm^{-2}s^{-1}keV^{-1}}$ is
given by
\begin{equation}
 \frac{dN}{dE} = 9.412\times10^{-3}\left(\frac{E}{{\rm
1~keV}}\right)^{-1.29}\exp{\left(-\frac{E}{{\rm 40~keV}}\right)}. 
\label{eq1}
\end{equation}

 The 20--50~keV energy flux of the CXB model (Eq.~\ref{eq1}),
  $5.8\times10^{-8}~{\rm erg\,s^{-1}\,cm^{-2}\,sr^{-1}}$, agrees with
  the recent report from the {\it Beppo-SAX} observations
  \citep{2007ApJ...666...86F} within about $8$\%. This level of
  uncertainty in the CXB model is smaller than the systematic error of
  the current NXB model. However, according to the calibration report
  based on the Crab observations, the PIN spectral normalization is
  systematically higher than the previous results given in
  \cite{1974AJ.....79..995T} by 13\%. Thus, we increase CXB given in
  Eq.~\ref{eq1} by a factor of 1.13 in the simulation.

The resulting simulated CXB spectrum is shown in Fig.~\ref{fig7}. To
calculate the CXB spectrum the integration time of 100~Ms was assumed;
thus, the statistical error of the simulated spectrum is negligibly
small. The CXB count rates in the 12--40~keV and 40--60~keV bands are
given by $2.9\times10^{-2}$ and $6.1\times10^{-4}~{\rm c\,s^{-1}}$,
respectively, which amount to about 6.1\% and 1.2\% of the NXB,
respectively.

\cite{2002PASJ...54..327K} have found that, based upon the {\it ASCA}
observations, the CXB fluctuation in the 2--10~keV band is $\sim 6$\%,
which can be attributed to the Poisson noise of the source count
within $0.5~{\rm degree^2}$. Moreover, by scaling the {\it HEAO-1}
result of 2.8\% with the equation $\sigma_{\rm CXB}\propto
\Omega^{-0.5}S^{0.25}$\citep[$\Omega$ and $S$ are the effective
  solid angle of the observation and upper cut-off flux of detectable
  discrete sources in the field of view,
  respectively;][]{1974ApJ...188..279C}, \cite{2008arXiv0805.3582K}
calculated the $1\sigma$ fluctuation of CXB to be 9.2\% for the
$\sim100$~ks HXD/PIN observations.  Therefore, the CXB fluctuation
over the $30\arcmin$ scale is small compared to the NXB systematic
error.

\subsection{Hard X-ray point sources}\label{subsec:pin_source}
Next, we estimate the hard X-ray flux of point sources inside the
field of view of HXD/PIN from the $30$~ks {\it XMM-Newton}/PN data on
the same field (Observation ID: 0112960101).  The data reduction was
carried out in the standard manner with {\tt SAS} version 7.1.0 and
the periods of background flares were removed with the threshold value
of $0.22~{\rm c\,s^{-1}}$ in the 12--14~keV band.

With the {\tt edetect\_chain} program, thirty point sources have been
detected in the PN image. The photon index for each source was
estimated from the hardness ratio, defined as a ratio of the source
count rates in the 0.5--1.5~keV and 1.5--5~keV band.  The background
was subtracted using the blank-sky data
\citep{2007A&A...464.1155C}. The galactic absorption was assumed in
the calculation.  The photon index of $\Gamma\gtrsim 2$ was obtained
for almost all the sources that have significant emission in the hard
band. We then simulate the PIN spectrum expected for each source using
a power-law spectrum with $\Gamma=2$, where we include the PIN angular
response function using {\tt hxdarfgen}.

Fig.~\ref{fig7} shows the spectrum of the sum of the detected
sources. We find that the contribution from the point sources is
totally negligible.

\subsection{PIN source spectrum}\label{subsec:pin_1apec}
%%%%%%%%%%%%%%%%%%%%%%%%%%%%%%%%%%%%%%%%%%%%
\begin{figure}
\centering
\rotatebox{270}{\scalebox{0.33}{\includegraphics{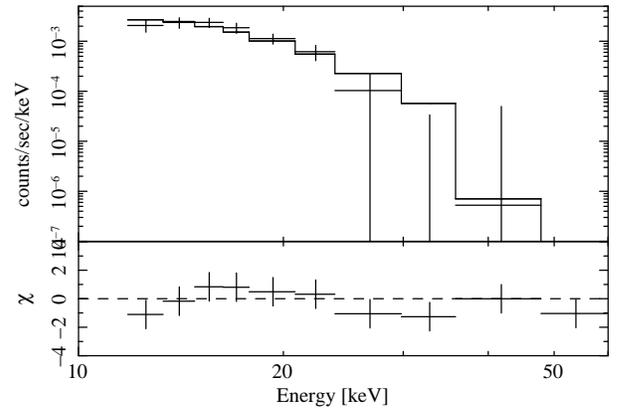}}}
\caption{Observed PIN spectrum with the background model subtracted
  (Data$-$NXB$-$CXB; the crosses) in the 12--60~keV band. The
  error-bars show only the statistical errors. The histogram shows the
  best-fitting APEC thermal plasma model. The bottom panel shows the
  residuals in units of the number of standard deviations.}
\label{fig8}
\end{figure}
%%%%%%%%%%%%%%%%%%%%%%%%%%%%%%%%%%%%%%%%%%%%
%%%%%%%%%%%%%%%%%%%%%%%%%%%%%%%%%%%%%%%%%%%%
\begin{table*}
\centering
\caption{Single-temperature APEC model parameters fit to the PIN data in
  the 12--60~keV band. The systematic error in the NXB model is
  included in the error estimate.}
\label{tab6}
\begin{tabular}{llllll} \hline\hline 
Sensor & $kT$~[keV] & $Z$~[solar] & $K$
  & $\chi^2/{\rm dof}$ & NXB$^{\mathrm{a}}$ \\ \hline
PIN & 20.2 (12.2 -- 41.6) & 0.33(Fix) & 1.02(0.59--1.97)$\times10^{-2}$ & 6.7/8 & $1.00$\\ 
PIN & 11.5 & 0.33(Fix) & 0.76 $\times10^{-2}$ & 26.2/8 & $1.03$  \\ 
PIN & 27.9   & 0.33(Fix) & 1.28$\times10^{-2}$ & 6.9/8 & $0.97$ \\ \hline
\end{tabular}
\begin{list}{}{}
\item[$^{\mathrm{a}}$] The NXB normalization factor.
\end{list}
\end{table*}
%%%%%%%%%%%%%%%%%%%%%%%%%%%%%%%%%%%%%%%%%%%%

In Fig.~\ref{fig8} the observed PIN spectrum with NXB and CXB
subtracted is shown.  In the 12--40~keV band we detect the hard X-ray
emission at the $9\sigma$ level. Taking into account the NXB
systematic error, however, significance of the detection becomes much
lower: the estimated count rate is $0.0187\pm 0.0021\,(\pm
  0.0142)~{\rm c\,s^{-1}}$ in the 12--40~keV, where the first and
second errors are the $1\sigma$ statistical and $1\sigma$ systematic
uncertainties, respectively.  At higher energies, 40--60~keV, the
count rate is $7\pm6\,(\pm15) \times 10^{-4}~{\rm c\,s^{-1}}$ ($1
\sigma$); thus, we do not expect significant hard X-ray emission in
the 40--60~keV band.

Next, we fit the single-temperature APEC model to the
background-subtracted PIN spectrum in the 12--60~keV band. (The XIS
data is not used here. For a joint analysis of XIS and HXD/PIN, see
\S~\ref{sec:xis+hxd_analysis}.)  The metal abundance was fixed at the
best-fitting value obtained from the XIS analysis, 0.33~solar.  The
systematic error of the spectral parameters due to the systematic
uncertainty in the NXB model was estimated by changing the
normalization factor of the NXB model by $\pm 3$\% and repeating the
fitting procedure.

The results are shown in Fig.~\ref{fig8} and Table~\ref{tab6}.  We
find that the temperature and normalization of the single-temperature
APEC model are $kT=20.2^{+21.4}_{-8.0}\,(^{+7.7}_{-8.7})$~keV and
$1.02^{+0.95}_{-0.43}\,(^{+0.26}_{-0.26}) \times 10^{-2}$,
respectively. These parameters are consistent with those obtained from
the XIS data. Therefore, we conclude that the hard X-ray emission
observed with PIN does indeed come from \object{RX~J1347.5--1145}.

From the observed PIN count rate and the best-fitting
single-temperature APEC model given in Table~\ref{tab2}, we estimate
the energy flux to be $3.8 \pm 0.5\,(\pm 3.3) \times 10^{-12}~{\rm
  erg\,s^{-1}cm^{-2}}$ in the 12--60~keV ($1\sigma$).  On the other
hand, the previous {\it Beppo-SAX} observation of
\object{RX~J1347.5--1145} has reported detection of the hard X-ray
emission at the $\sim1.5\sigma$ significance:
$(4.0\pm2.6)\times10^{-2}~{\rm c\,s^{-1}}$ in the 13--60~keV band
\citep{2001MNRAS.322..187E}, and $(6.8\pm4.4)\times10^{-2}~{\rm
  c\,s^{-1}}$ in the 20--80~keV band \citep{2004ApJ...608..166N}. With
the 12.9~keV Raymond-Smith model and the response of the PDS
instrument on board {\it Beppo-SAX}, the {\tt pimms} program gives the
12--60~keV flux as $(4.5\pm2.9)\times 10^{-12}~{\rm erg\,s\,cm^{-1}}$.
(All errors quoted here are $1\sigma$.)  Thus, the PIN result agrees
with that of PDS to within the $1\sigma$ errors.

\section{XIS+HXD joint analysis}\label{sec:xis+hxd_analysis}
In \S~\ref{sec:xis_analysis} the {\it Suzaku} XIS data indicated the
presence of a significant amount of very hot gas. In this section, we
study the physical properties of this hot component, such as the
amount, temperature, and metal abundance, by performing a joint
analysis of the {\it Suzaku} data and the spatially-resolved {\it
  Chandra} spectra.

In \S~\ref{subsec:deproject} the properties of the ambient
gas are derived from the {\it Chandra} data.  In \S~\ref{subsec:xis+hxd_se} 
the properties of the hot gas component in the SE quadrant are derived from
the {\it Suzaku} broad-band spectra combined with the {\it Chandra}
data by means of a multi-temperature model.  In
\S~\ref{subsec:xis+hxd_nonthermal} we argue that the hot component is
explained better by thermal gas than by non-thermal gas, and derive an
upper limit on the possible non-thermal emission.

\subsection{Modeling the cluster average component with {\it
    Chandra}}\label{subsec:deproject}
%%%%%%%%%%%%%%%%%%%%%%%%%%%%%%%%%%%%%%%%%%%%
\begin{figure*}
\centering
\rotatebox{270}{\scalebox{0.5}{\includegraphics{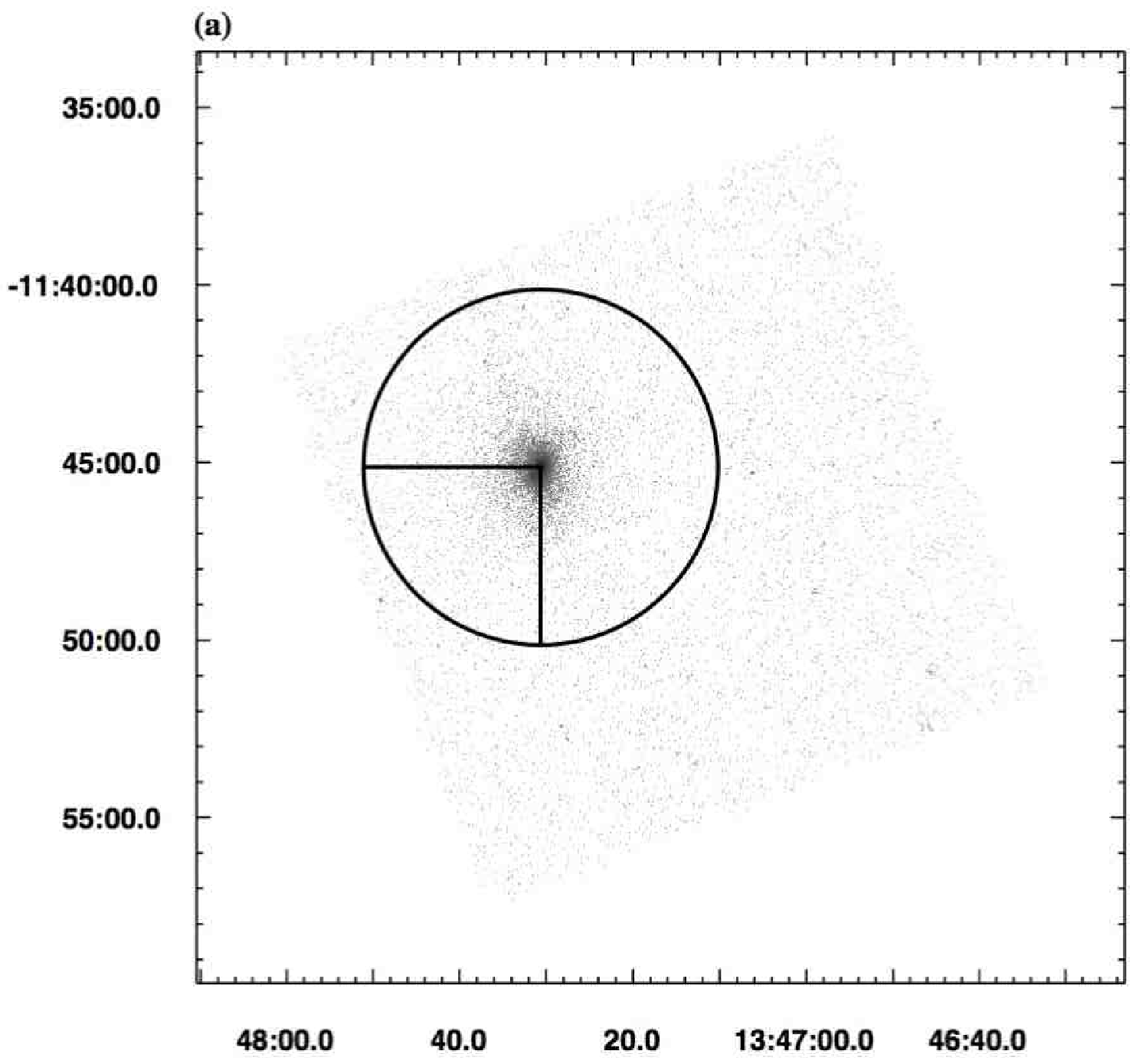}}}
\rotatebox{270}{\scalebox{0.33}{\includegraphics{f9b.eps}}}
\caption{(a) {\it Chandra} ACIS-I image in the 0.5--7~keV band. The
  solid lines show the NW and SE regions. The radius of the circle is
  5\arcmin.  (b) The surface brightness profiles of the position angle
  $0^{\circ}-360^{\circ}$ (black), NW:$-90^{\circ}-180^{\circ}$ (red),
  and SE:$180^{\circ}-270^{\circ}$ (green).  The black line shows the
  best-fitting double $\beta$-model for the position angle of
  $0^{\circ}-360^{\circ}$.}
\label{fig9}
\end{figure*}
%%%%%%%%%%%%%%%%%%%%%%%%%%%%%%%%%%%%%%%%%%%%
%%%%%%%%%%%%%%%%%%%%%%%%%%%%%%%%%%%%%%%%%%%%
\begin{figure}
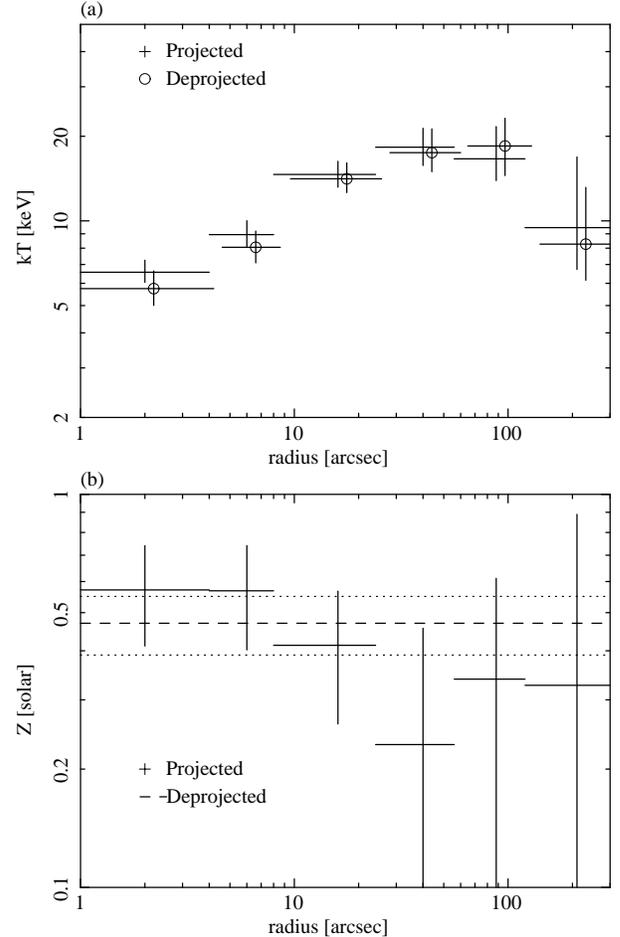

\centering
\rotatebox{270}{\scalebox{0.33}{\includegraphics{f10a.eps}}}
\rotatebox{270}{\scalebox{0.33}{\includegraphics{f10b.eps}}}
\caption{(a) Projected (crosses) and deprojected (circles) temperature
  profiles measured from the {\it Chandra} data in the NW region.  (b)
  Projected metal abundance profile (crosses). The horizontal dashed
  and dotted lines show the best-fitting constant abundance and the
  corresponding 90\% confidence interval obtained from the
  deprojection analysis, respectively. We shall assume that the metal
  abundance is constant over all radial bins throughout this paper. }
\label{fig10}
\end{figure}
%%%%%%%%%%%%%%%%%%%%%%%%%%%%%%%%%%%%%%%%%%%%
%%%%%%%%%%%%%%%%%%%%%%%%%%%%%%%%%%%%%%%%%%%%
\begin{table*}
\centering
\caption{Parameters of the ``6APEC model,'' estimates of the deprojected
  temperature profile from the {\it Chandra} ACIS-I data at the six
  annular bins in the NW region, which excludes the SE quadrant.}
\label{tab7}
\begin{tabular}{lllll} \hline\hline 
Radius~[\arcsec]       & $kT$~[keV] & $Z$~[solar]$^{\mathrm{a}}$ & $K$$^{\mathrm{b}}$ & $\chi^2/{\rm dof}$\\ \hline
0--4       &5.74 (5.01 -- 6.64)    & 0.47 (0.39 -- 0.55)& $0.97 (0.92 - 1.02)\times10^{-3}$ & 364.2/290 \\
4--8     &8.06 (7.10 -- 9.20)    & 	 	   & $1.83 (1.77 - 1.89)\times10^{-3}$ & \\
8--24    &14.12 (12.61 -- 16.08) & 		   & $4.02 (3.93 - 4.10)\times10^{-3}$ & \\
24--56   &17.49 (14.95 -- 21.26) & 		   & $2.86 (2.79 - 2.95)\times10^{-3}$ & \\
56--120  &18.47 (14.49 -- 23.18) & 		   & $2.29 (2.22 - 2.37)\times10^{-3}$ & \\
120--300 &8.27 (6.15 -- 13.16)   &   		   & $0.91 (0.86 - 0.97)\times10^{-3}$ & \\ \hline
\end{tabular}
\begin{list}{}{}
\item[$^{\mathrm{a}}$] The metal abundance is assumed to be common for all radial bins. 
\item[$^{\mathrm{b}}$] $K$ represents the normalization factor for
  each spherical shell, i.e., the position angle of
  $0^{\circ}-360^{\circ}$, of the fitting.
\end{list}
\end{table*}
%%%%%%%%%%%%%%%%%%%%%%%%%%%%%%%%%%%%%%%%%%%%
The deep {\it Chandra} ACIS-I data on \object{RX~J1347.5--1145} (Obs
ID:3592, Date:2003-09-03) was analyzed with {\tt CIAO} version 3.4 and
{\tt CALDB} version 3.3.0.  The net exposure time after removing the
periods of high background rates is 56.1~ks. The backgrounds were
estimated from the same detector regions of the blank-sky data, whose
normalization factors were determined with the ratios of the
10--12~keV count rates.

Fig.~\ref{fig9} shows the {\it Chandra} image and the
azimuthally-averaged surface brightness profiles for three different
position angles ($0^{\circ}-360^{\circ}$,
NW:$-90^{\circ}-180^{\circ}$, and SE:$180^{\circ}-270^{\circ}$).  The
cluster emission extends out to $r=5\arcmin \sim 1.7~{\rm Mpc}$ and
the significant excess emission is visible in $10\arcsec \lesssim
r\lesssim 60\arcsec$ for the SE quadrant.

To derive the average temperature profile within $5\arcmin$, we
measure the spectra in the following six radial bins in the NW region,
i.e., the region outside of the SE quadrant: $0-4\arcsec$,
$4\arcsec-8\arcsec$, $8\arcsec-24\arcsec$, $24\arcsec-56\arcsec$,
$56\arcsec-120\arcsec$, and $120\arcsec-300\arcsec$.  The radial bins
were chosen such that the temperature profile is measured with
sufficient spatial resolution and the statistical accuracy $\sim
20$\%.

Fig.~\ref{fig10} shows the projected temperature profile, i.e., the
temperature profile determined from the simple APEC model fit to each
radial bin in the NW region. Our estimate of the projected temperature
profile derived from the {\it Chandra} ACIS-I data agrees with the
previous {\it Chandra} ACIS-S results reported by
\cite{2002MNRAS.335..256A} to within the statistical error,
particularly in the central 100\arcsec region.  We have also confirmed
based on the ACIS-I spectral simulations assuming the same photon
statistics as the present observation that we can securely determine
the temperature of each bin including the high-temperature ($\sim
18$~keV) regions between 24\arcsec and 120\arcsec.

Since each spectrum is a superposition of the cluster emission from
any points along the line of sight, we correct this effect, or
``deproject,'' by using the PROJCT model in the {\tt XSPEC} software
and fitting the six radial bins simultaneously under the assumption
that the temperature distribution is spherically symmetric. As noted
in \cite{2002MNRAS.335..256A}, the metal abundance gradient is
marginally seen in the {\it Chandra} data; however, the large
uncertainty in the metal abundance measurement does not warrant our
treating it as a free function. We therefore assume that the metal
abundance distribution is uniform and ignore the abundance gradient.

The resulting deprojected temperature profile in the NW region is
shown in Fig.~\ref{fig10} and the best-fitting parameters are given in
Table~\ref{tab7}. Henceforth, we refer to the best-fitting model
presented in Table~\ref{tab7} as the ``6APEC model.''

\subsection{Temperature measurement of the SE clump}\label{subsec:xis+hxd_se} 
%%%%%%%%%%%%%%%%%%%%%%%%%%%%%%%%%%%%%%%%%%%%
\begin{figure}
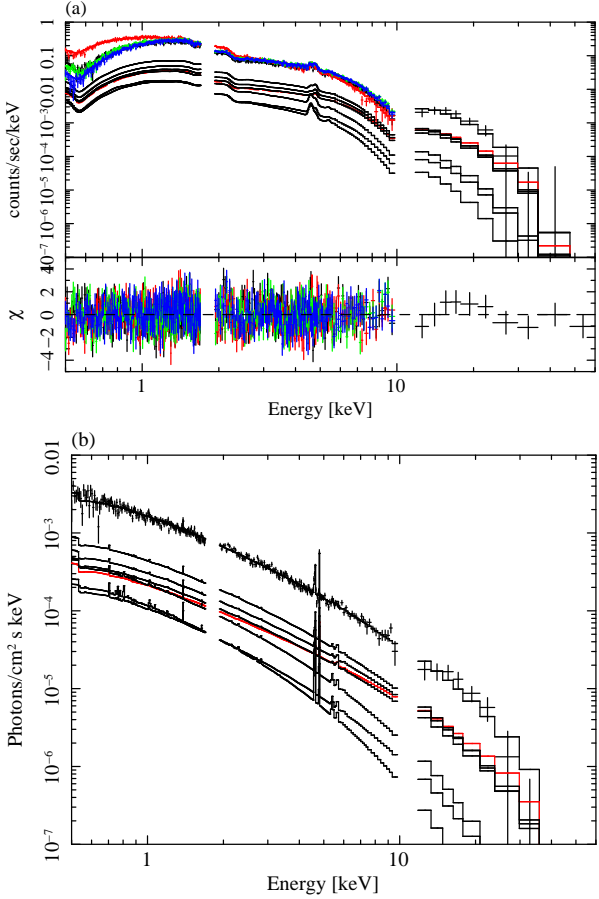

\centering
\rotatebox{270}{\scalebox{0.33}{\includegraphics{f11a.eps}}}
\rotatebox{270}{\scalebox{0.33}{\includegraphics{f11b.eps}}}
\caption{(a) 6APEC$\times$PLABS($\alpha=0.02$) + APEC model, fit
  simultaneously to the observed XIS (0.5--10~keV) and PIN
  (12--60~keV) spectra. The spectra taken by four sensors, XIS-0
  (black), 1 (red), 2 (green), and 3 (blue), are shown separately.
  The black and red lines show the ambient component
  (6APEC$\times$PLABS) and the excess hot component described by an
  additional APEC model, respectively, convolved with the telescope
  and the detector response functions.  In the bottom panel the
  residuals of the fit in units of the number of standard deviations
  are shown.  We have used the nominal NXB model for the background of
  the PIN data (see \S~\ref{subsec:pin_nxb}).  (b) same as (a), but
  for the unfolded XIS-0 and PIN spectra. }
\label{fig11}
\end{figure}
%%%%%%%%%%%%%%%%%%%%%%%%%%%%%%%%%%%%%%%%%%%%
%%%%%%%%%%%%%%%%%%%%%%%%%%%%%%%%%%%%%%%%%%%%
\begin{table*}
\centering
\caption{Spectral parameters for the SE excess component obtained
  from the 6APEC$\times$PLABS+APEC model fit to the XIS+PIN data and the XIS data.}
\label{tab8}
\begin{tabular}{lllllll} \hline\hline 
       & \multicolumn{3}{c}{APEC} & & \\ \cline{2-4}
Sensor & $kT_{\rm ex}$~[keV] & $Z_{\rm ex}$~[solar]$^{\mathrm{a}}$ & $K_{\rm ex}$
  & $\chi^2/{\rm dof}$ & $\alpha$$^{\mathrm{b}}$\\ \hline
XIS0--3+PIN & 25.3(20.8--31.4) & 0.38(0.35--0.41) & 
2.14(2.05--2.24)$\times10^{-3}$ & 1311.1/1219 & 0.02 \\
XIS0--3+PIN & 15.8 & 0.36 & 1.98$\times10^{-3}$ & 1313.1/1219 & 0.00 \\ 
XIS0--3+PIN & 32.2 & 0.39 & 2.26$\times10^{-3}$ & 1310.2/1219 & 0.03 \\\hline
XIS0--3 & 24.8(20.2--30.9) & 0.38(0.35--0.41) & 
 2.00(1.91--2.09)$\times10^{-3}$ & 1303.3/1209 & 0.02 \\ 
XIS0--3 & 15.5 & 0.36 & 1.84$\times10^{-3}$ & 1303.7/1209 & 0.00 \\
XIS0--3 & 31.9 & 0.39 & 2.11$\times10^{-3}$ & 1303.0/1209 & 0.03 \\ \hline
\end{tabular}
\begin{list}{}{}
\item[$^{\mathrm{a}}$] The metal abundance is assumed to be the same
  for all the APEC components, including the ambient gas as well as
  the excess component.
\item[$^{\mathrm{b}}$] The power-law index of the PLABS model, coming
from the {\it Suzaku}-{\it Chandra} cross-calibration (see text).
\end{list}
\end{table*}
%%%%%%%%%%%%%%%%%%%%%%%%%%%%%%%%%%%%%%%%%%%%
The 6APEC model in \S~\ref{subsec:deproject} describes the average
temperature structure of ambient gas, measured from the NW region. We
subtract the flux accounted by this model from the {\it Suzaku} data,
and study the nature of the excess emission in the SE quadrant.

We add another APEC model to describe the hot component in the SE
quadrant. We use the best-fitting parameters for $kT$ and $K$ given in
Table~\ref{tab7} for the ambient gas (6APEC model), and use the {\it
  Suzaku} data, both the XIS and PIN spectra, to constrain the
temperature, $kT_{\rm ex}$, and the normalization, $K_{\rm ex}$, of
the excess component in the SE quadrant. As for the metal abundance,
the abundance is assumed to be equal to all of the six radial bins of
the ambient gas, as well as to the excess component, i.e., a single
abundance is applied to the entire cluster. We do this because the
metal abundance is not constrained by the {\it Chandra} data very
well.

We have found that the {\it Chandra} data tend to give systematically
higher values for $kT$ and $K$ than the XIS data do. In order to
correct for the difference between their instrumental calibrations, we
multiply the 6APEC model by the PLABS model, $M(E)=E^{-\alpha}$, with
$\alpha=0.02$ ($E$ is the energy of photons), and use the relative
normalization factor of 0.93. See
Appendix~\ref{appendix:suzaku_chandra} for more details on the {\it
  Suzaku}-{\it Chandra} cross-calibration.

Fig.~\ref{fig11} and Table~\ref{tab8} show the results of the
6APEC$\times$PLABS + APEC model fit to the XIS (0.5--10~keV) and PIN
(12--60~keV) data. To show the degree of systematic errors due to the
uncertainty in the {\it Suzaku}-{\it Chandra} cross-calibration, we
also show the parameters for $\alpha=0$ and 0.03 in Table~\ref{tab8}.
From these results we conclude that the temperature of the excess
emission in the SE quadrant is $kT_{\rm ex}=25.3^{+6.1}_{-4.5}
\,(^{+6.9}_{-9.5})$~keV, where the first error is statistical and the
latter is systematic (both 90\% C.L.).  In Table~\ref{tab8}, we also
show the fitted parameters when only XIS was used.  Thanks to the high
photon statistics of XIS, the temperature of the SE excess component
can be constrained as $kT_{\rm ex}^{\rm XIS}=24.8^{+6.1}_{-4.6}\,
(^{+6.8}_{-9.5})$~keV. However, the spectral cut-off at $25/(1+z)\sim
17$~keV actually falls into the PIN band, which suggests the
importance of the broad-band spectroscopy in the present study.  As
for the other quantities, we find from the XIS+PIN joint analysis that
the metal abundance is 0.38(0.35--0.41)~solar, and the
absorption-corrected X-ray luminosity of the excess component is
$L_{\rm X,ex}= 1.4\times10^{45}~{\rm erg\,s^{-1}}$ in the 0.5--10~keV,
and $1.2\times10^{45}~{\rm erg\,s^{-1}}$ in the 12--60~keV.

Since the determination of $kT_{\rm ex}$ from the X-ray spectroscopy
is the central result of this paper, we estimate the other potential
source of systematic errors in our analysis. Specifically, we examine:
(i) the assumption of constant metal abundance over all radii, (ii)
some arbitrariness in the choice of the SE region, and (iii) effects
of the uncertainty in the 6APEC model parameters.

\begin{itemize}
\item[(i)] What if the metal abundance of the ambient gas and that of
  the excess component are different? What if the abundance gradient
  does exist in the ambient gas? To explore these possibilities we
  treat the metal abundance of the excess component, $Z_{\rm ex}$, as
  well as that of the ambient gas in two larger radial bins,
  $0\arcsec<r<8\arcsec$ and $8\arcsec< r<300\arcsec$, as free
  parameters, and fit them simultaneously. We find $kT_{\rm
    ex}=26.3$~keV, $Z_{\rm ex}<1.8$~solar, and $K_{\rm
    ex}=2.16\times10^{-3}$ ($\chi^2/{\rm dof}=1306/1217$), which are
  consistent with the values given in Table~\ref{tab8}. As for the
  metal abundance of the ambient gas we find $Z\sim0.64$~solar and
  $0.22$~solar in $0\arcsec<r<8\arcsec$ and $8\arcsec< r<300\arcsec$,
  respectively.
\item[(ii)] From the {\it Chandra} image one may observe that the excess
  emission appears to extend over slightly more than the SE
  quadrant. How does this affect the determination of $kT_{\rm ex}$?
  We repeat our analysis with the SE and NW regions re-defined as
  $(180^{\circ}-315^{\circ})$ and $(-45^{\circ}-180^{\circ})$,
  respectively. We find $kT_{\rm ex}=20.0$~keV, $K_{\rm
    ex}=2.87\times10^{-3}$, and $Z_{\rm ex}=Z=0.37$~solar with
  $\chi^2/{\rm dof}=1315/1219$. Therefore, the exact choice of the SE
  region does not change the best-fitting parameters very much.
\item[(iii)] The statistical errors in the 6APEC model parameters had
  not been propagated through the spectral fitting of the SE quadrant,
  except for the cross-calibration error in the spectral slope, which
  was included by varying $\alpha$ from 0 to 0.03. To address this, we
  carry out simultaneous APEC fits to the {\it Chandra} spectra in 6
  radial bins (without deprojection) as well as to the {\it Suzaku}
  XIS and PIN spectra, for which we use the 6 APEC models multiplied
  by the $\alpha=0.02$ PLABS model plus the additional APEC
  model. We find $kT_{\rm ex}>26$~keV, $Z_{\rm
      ex}=0.38(0.35-0.41)$~solar, and $K_{\rm
      ex}=2.16(1.96-2.41)\times10^{-3}$ ($\chi^2/{\rm
      dof}=1676/1510$).  Although the temperature is not well
    constrained in this case, it equally shows a high value. 
\end{itemize}
From these studies we conclude that our estimate, $kT_{\rm
  ex}=25.3^{+6.1}_{-4.5}\,(^{+6.9}_{-9.5})$~keV, is robust.

Finally, as a consistency check we analyze the {\it Chandra} spectrum
of the SE quadrant ($r<5'$) on its own, without the {\it Suzaku} data,
in the same manner. The best-fitting parameters for the excess
component are $kT_{\rm ex}^{Chandra}=31.1^{+24.1}_{-12.6}$~keV,
$Z_{\rm ex}^{Chandra}=0.49^{+0.15}_{-0.15}$~solar, and $K_{\rm
  ex}^{Chandra}=1.47^{+0.21}_{-0.15}\times10^{-3}$ ($\chi^2/{\rm
  dof}=320/332$). Therefore, both the {\it Chandra}-alone analysis and
the joint {\it Chandra}+{\it Suzaku} analysis show that the excess
emission has the temperature in excess of 20~keV.\footnote{As for the
  normalization factor, $K_{\rm ex}^{Chandra}$ is found to be smaller
  than that from the joint analysis by about 35\%. This may be
  explained by the difference in their integration regions. On the
  other hand, this level of difference actually amounts only to a
  few\% of the total cluster emission, and is comparable to the
  calibration errors of the instruments. Therefore, we simply assign
  30\% systematic error to $K_{\rm ex}$. }

Compared to the results from the {\it Chandra} data alone, the joint
{\it Chandra}+{\it Suzaku} broad-band data analysis yields much more
accurate determination of $kT_{\rm ex}$. The {\it Suzaku}'s
unprecedented sensitivity over the wide X-ray band makes it possible to
determine, for the first time, the temperature of such hot gas in the
ICM solely from the X-ray spectroscopy without help of the SZ effect.

\subsection{Constraint on the non-thermal emission}\label{subsec:xis+hxd_nonthermal}
%%%%%%%%%%%%%%%%%%%%%%%%%%%%%%%%%%%%%%%%%%%%
\begin{figure}
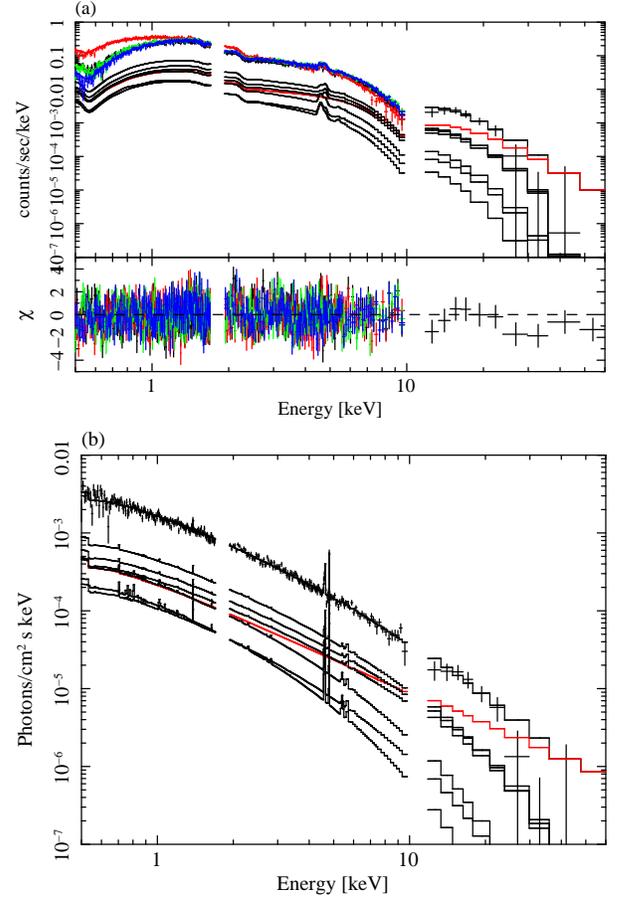

\centering
\rotatebox{270}{\scalebox{0.33}{\includegraphics{f12a.eps}}}
\rotatebox{270}{\scalebox{0.33}{\includegraphics{f12b.eps}}}
\caption{(a) 6APEC$\times$PLABS($\alpha=0.02$) (thermal model for the
  ambient gas) + PL model (non-thermal model for the hot gas), fit
  simultaneously to the observed XIS (0.5--10~keV) and PIN
  (12--60~keV) spectra. The spectra taken by four sensors, XIS-0
  (black), 1 (red), 2 (green), and 3 (blue), are shown separately.
  The black and red lines show the ambient component
  (6APEC$\times$PLABS) and the excess hot component described by a
  non-thermal power-law model, respectively, convolved with the
  telescope and the detector response functions.  In the bottom panel
  the residuals of the fit in units of the number of standard
  deviations are shown.  We have used the nominal NXB model for the
  background of the PIN data (see \S~\ref{subsec:pin_nxb}).  (b) same
  as (a), but for the unfolded XIS-0 and PIN spectra.}
\label{fig12}
\end{figure}
%%%%%%%%%%%%%%%%%%%%%%%%%%%%%%%%%%%%%%%%%%%%
%%%%%%%%%%%%%%%%%%%%%%%%%%%%%%%%%%%%%%%%%%%%
\begin{table*}
\centering
\caption{Spectral parameters for the SE excess component obtained
  from the 6APEC$\times$PLABS+non-thermal power-law (PL) model fit to the XIS+PIN data.}
\label{tab9}
\begin{tabular}{lllllll} \hline\hline 
            & \multicolumn{2}{c}{Power-law} & 6APEC & \\ \cline{2-3}
Sensor      & $\Gamma$ & Normalization  & $Z$~[solar] & $\chi^2/{\rm dof}$ & $\alpha$\\ \hline
XIS0--3+PIN & 1.45(1.41--1.48) & 2.58(2.30--2.52)$\times10^{-4}$ &0.42(0.39--0.45) & 1317.1/1219 & 0.02 \\ 
XIS0--3+PIN & 1.54 & 2.57$\times10^{-4}$ & 0.40 & 1312.5/1219 & 0.00 \\ 
XIS0--3+PIN & 1.41 & 2.58$\times10^{-4}$ & 0.42 & 1321./1219 & 0.03 \\ \hline
\end{tabular}
\end{table*}
%%%%%%%%%%%%%%%%%%%%%%%%%%%%%%%%%%%%%%%%%%%%
%%%%%%%%%%%%%%%%%%%%%%%%%%%%%%%%%%%%%%%%%%%%
\begin{table*}
\centering
\caption{Limits on the amplitude (normalization) of a non-thermal
  power-law (PL) component with $\Gamma=1.5$ as the explanation of the
  excess hard X-ray emission in the SE quadrant. }
\label{tab10}
\begin{tabular}{llllllll} \hline\hline 
Sensor  & Model$^{\mathrm{a}}$  & \multicolumn{2}{c}{Normalization$^{\mathrm{b}}$} & \multicolumn{2}{c}{$\chi^2/{\rm dof}$$^{\mathrm{c}}$} & Flux(12--60~keV)$^{\mathrm{d}}$ \\ \cline{3-4} \cline{5-6}
& & NXB$\times1.00$ & NXB$\times0.97$ & NXB$\times1.00$ & NXB$\times0.97$ & [${\rm erg\,s^{-1}\,cm^{-2}}$] \\ \hline
PIN         & APEC+PL & $\sim 0           $ & $5.1\times10^{-4}$  & 7.2/7 & 4.6/7 & $<2.11\times10^{-11}$ \\
XIS0--3+PIN & APEC+PL  & $<1.6\times10^{-4}$ & $2.0\times10^{-4}$  & 1310.2/1218 & 1344.3/1218 & $<8.2\times10^{-12}$ \\ \hline
PIN         & PL         & $1.8(0.7-2.9)\times10^{-4}$ & $6.3\times10^{-4}$  & 8.8/9 & 5.1/9 & $<2.15\times10^{-11}$ \\
XIS0--3+PIN & PL & $2.48(2.39-2.59)\times10^{-4}$ & $2.52\times10^{-4}$ & 1322.8/1220 & 1351.9/1220 & $<3.8\times10^{-12}$ \\\hline
\end{tabular}
\begin{list}{}{}
\item[$^{\mathrm{a}}$] The spectral model for the SE excess emission. 
\item[$^{\mathrm{b}}$] The amplitude (normalization) of the power-law
  component after removing the NXB models with rescaling factors of 1
  and 0.97. The error and limits are the $1\sigma$ bound.
\item[$^{\mathrm{c}}$] The values of $\chi^2$ and the number of
  degrees of freedom.
\item[$^{\mathrm{d}}$] The $3\sigma$ upper limit on the energy flux of
  the power-law component in the 12--60~keV band.
\end{list}
\end{table*}
%%%%%%%%%%%%%%%%%%%%%%%%%%%%%%%%%%%%%%%%%%%%

So far, we have been assuming that the excess hard X-ray emission from
the SE quadrant is thermal; however, could it actually be non-thermal?
If so, our derived temperature must be reconsidered.

To address this question, the XIS and PIN data is re-analyzed in light
of non-thermal emission. While the 6APEC thermal plasma model is again
used for the ambient gas, the APEC model is replaced with a power-law
(PL) spectrum for describing the excess component. We then fit the new
model, ``6APEC$\times$ PLABS+PL model,'' to the observed XIS and PIN
data.  The amplitude (normalization), a power-law index, $\Gamma$, of
the non-thermal component, and the metal abundance of the ambient gas
are treated as free parameters.

Fig.~\ref{fig12} shows the results, and Table~\ref{tab9} lists the
best-fitting parameters. We find that the power-law index is given by
$\Gamma\sim 1.5$, or more precisely
$\Gamma=1.45^{+0.03}_{-0.04}\,(^{+0.09}_{-0.04})$.  The reduced
  $\chi^2$ of this model, $\chi^2=1317.1/1219$ (see Table~\ref{tab9}),
  is slightly larger than that for the thermal model, $\chi^2=
  1315.1/1219$ (see Table~\ref{tab8}), and negative residuals are seen
  in consecutive bins above 22~keV.

Let us examine the non-thermal model more closely.  In the XIS band
below 10~keV, the non-thermal model with $\Gamma=1.5$ and the thermal
APEC model with $25$~keV are hardly distinguishable.  However, the
observed PIN spectrum is much {\it softer} than $\Gamma=1.5$, with the
effective photon index being $\Gamma_{\rm eff}\sim 3$.  When we fit
the 6APEC$\times$ PLABS+PL model to the PIN data alone, the 90\% lower
bound on the photon index is found as $\Gamma>1.8$\footnote{The
  systematic errors due to both the NXB model and the {\it
    Suzaku}-{\it Chandra} cross-calibration are included.}, which is
significantly above the 90\% upper bound on $\Gamma$ from the XIS
data, $\Gamma<1.5$. We have further checked that fixing the power-law
index to $\Gamma=2.0$, as predicted from the non-thermal
bremsstrahlung in the strong shock limit \citep{2000ApJ...533...73S},
yields a poor fit to the XIS and PIN data; the reduced $\chi^2$ is
1708 for 1220 degrees of freedom. Therefore, it seems difficult to
explain the SE excess component with the non-thermal PL model.

Given that the excess component is explained better by the thermal
emission than by the non-thermal emission, we derive an upper limit on
the non-thermal emission as the explanation for the SE component. We
fix the metal abundance and the PL index at $0.38$~solar and
$\Gamma=1.5$, respectively.  The 6APEC model modified by the
$\alpha=0.02$ PLABS model is again used for describing the ambient
gas.

The derived upper limits are given in Table~\ref{tab10}. Taking into
account the NXB systematic error of the PIN detector, the $3\sigma$
upper limit on the non-thermal flux in the 12--60~keV, $F_{\rm HXR}$,
is estimated as $F_{\rm HXR}<2.1\times10^{-11}~{\rm
  erg\,s^{-1}cm^{-2}}$ from the PIN analysis, and $F_{\rm
  HXR}<8\times10^{-12}~{\rm erg\,s^{-1}cm^{-2}}$ from the XIS+PIN
simultaneous fit.  Here, $1\sigma$ error is calculated by adding the
$1\sigma$ statistical error and the $1\sigma$ systematic error in
quadrature.

\section{Discussion}\label{sec:discussion}
Using the deep broad-band observations of \object{RX~J1347.5--1145}
with {\it Suzaku}, the temperature structure has been determined with
a particular focus on the extremely hot gas.  We have found that the
multi-temperature thermal emission model including the very hot
($kT_{\rm ex}\sim 25$~keV) component fits the wide-band spectra
well. We have also placed an upper limit on the non-thermal hard X-ray
emission.

In \S~\ref{subsec:discussion_se} our results are compared with the
previous X-ray and radio observations, and examine the physical
properties of the SE clump further.  In \S~\ref{subsec:discussion_nt}
our results are compared with the radio mini halo in this cluster that
has been discovered from the 1.4~GHz radio observations to estimate
the magnetic field strength in the ICM.

\subsection{Properties of the extremely hot gas}\label{subsec:discussion_se}
From the analysis of the {\it Suzaku} spectra of
\object{RX~J1347.5--1145}, the temperature of the SE excess emission
is obtained to be $kT_{\rm
  ex}=25.3^{+6.1}_{-4.5}\,(^{+6.9}_{-9.5})$~keV (90\% C.L.; the first
error being statistical and the second systematic). This is in an
excellent agreement with the previous measurement by
\cite{2004PASJ...56...17K}, $28.5\pm7.3$~keV (68\%; statistical only),
from a joint analysis of the SZ data
\citep{1999ApJ...516L...1K,2001PASJ...53...57K} and the {\it Chandra}
ACIS-S3 data \citep{2002MNRAS.335..256A}.  Here, we compare our
X-ray-only results with their SZ+X-ray results.

We estimate the gas density and gas mass of the SE excess component
from the {\it Suzaku} and {\it Chandra} data. For simplicity, we
assume that the extremely hot gas is uniformly distributed within a
sphere of a radius $R=25\arcsec\sim144$~kpc (because the excess is
present in $10\arcsec< r < 60\arcsec$; Fig.~\ref{fig9}b). Using the
measured spectral normalization, $K_{\rm ex}$, which depends on $n_{e}
n_{\rm H}V$ (where $V=4\pi R^3/3$ is the volume of the gas and $n_{\rm
  H}=0.86n_e$), the electron density, $n_{\rm ex}$, and the gas mass,
$M_{\rm gas, ex}$, are obtained as:
\begin{eqnarray}
n_{e,\rm ex}
 & = & (1.6 \pm 0.2)\times10^{-2}
      \left(\frac{V_{\rm ex}}{3.6\times10^{71}~{\rm cm^3}}\right)^{-1/2}~{\rm cm^{-3}}, \\
M_{\rm gas,ex} 
& = & (5.6 \pm 0.8)\times10^{12}
       \left(\frac{V_{\rm ex}}{3.6\times10^{71}~{\rm cm^3}}\right)^{1/2}~{\rm M_{\sun}}.
\end{eqnarray}
Here, we have propagated the systematic error in $K_{\rm ex}$ through
the final results (\S~\ref{subsec:xis+hxd_se}).  These results may be
compared with those obtained in \citet{2004PASJ...56...17K}: $n_{\rm
  ex}= (1.49 \pm 0.59)\times10^{-2}~{\rm cm^{-3}}$ and $M_{\rm gas,ex}
\sim 2 \times10^{12}~{\rm M_{\sun}}$. While the inferred number
densities are in an excellent agreement, the gas mass from our
analysis is more than a factor of two greater than that from
\citet{2004PASJ...56...17K}; however, the uncertainty in the gas mass
estimate from \citet{2004PASJ...56...17K} is large enough for them to
be consistent.  (The uncertainty is large because they did not assume
spherical geometry for the excess component, and the line-of-sight
extension of the SE component was poorly constrained.)

How does the excess component compare with the rest of the cluster?
The average (ambient) temperature and gas density for
$10\arcsec<r<60\arcsec$, excluding the SE quadrant, are $kT \sim
15$~keV from the {\it Chandra} spectrum and $n_{e}\sim
6.6\times10^{-3}~{\rm cm^{-3}}$ from Eqs. 3--5 of
\cite{2004PASJ...56...17K}, respectively. Thus, the SE clump exhibits
the temperature and the density that are higher than the ambient gas
in the same radial bins by factors of 1.6 and 2.4, respectively.  This
means that the excess hot component is over-pressured, and such a
region is expected to be short-lived, $\sim0.5$~Gyr
\citep{1999ApJ...520..514T}.

As already discussed in \cite{2004PASJ...56...17K}, the gas properties
can be explained by a fairly recent (within the last 0.5~Gyr or so),
bullet-like high velocity ($\Delta v\sim 4500~{\rm km\,s^{-1}}$)
collision of two massive ($5\times10^{14}{\rm M_{\sun}}$) clusters.

Moreover, the heat energy of the SE clump is estimated to be $ E_{\rm
  th, ex} = (3/2) kT_{\rm ex}(n_{e,{\rm ex}} + n_{\rm H, ex}) V_{\rm
  ex}\sim 6\times10^{62}~{\rm erg}. $ This huge amount of energy
cannot be easily produced by a central source in the cluster: one
would need an AGN with $10^{46}~{\rm erg\,s^{-1}}$ for at least 1~Gyr
and put the energy into the SE region without any radiative loss. On
the other hand, cluster mergers, which are the most energetic events
in the Universe after the Big Bang, will most naturally explain this
high energy phenomenon. Therefore our results do support the merger
scenario, solely from the X-ray spectroscopic data without help of the
SZ data.
  
\subsection{Estimation of the magnetic field}\label{subsec:discussion_nt}
What physics can we learn from our upper limit on the non-thermal hard
X-ray emission from the {\it Suzaku} HXD/PIN?  The non-thermal
  hard X-ray emission is produced via the inverse Compton (IC)
  scattering of relativistic electrons off the Cosmic Microwave
  Background (CMB) photons, and the same population of electrons also
  produce the synchrotron radiation.  From the exact derivations by
  \cite{1970RvMP...42..237B}, equations for the synchrotron emission
  at the frequency $\nu_{\rm Syn}$ and the IC emission at $\nu_{\rm
    IC}$ are:
\begin{eqnarray}
\frac{dW_{\rm Syn}}{d\nu_{\rm Syn} dt} & =& \frac{4\pi N_0 e^3 B^{(p+1)/2}}{m_e c^2}
\left(\frac{3e}{4\pi m_e c }\right)^{-(p-1)/2}a(p)\nu_{\rm Syn}^{-(p-1)/2}, \label{eq4}\\
\frac{dW_{\rm IC}}{d\nu_{\rm IC} dt} &=& \frac{8\pi^2 r_0^2}{c^2}h^{-(p+3)/2}N_0(k_B T_{\rm CMB})^{(p+5)/2}F(p)\nu_{\rm IC}^{-(p-1)/2}.\label{eq5}
\end{eqnarray}
Here $N_0$ and $p$ are the normalization and the power-law index of
the electron distribution, $N(\gamma)=N_0 \gamma^{-p}$ ($\gamma$ is
the Lorentz factor of the electron). $h$ is the Planck constant,
$T_{\rm CMB}$ is the CMB temperature and $T_{\rm CMB}=2.73(1+z)$~K. The functions $a(p)$
and $F(p)$ are defined as follows:
\begin{eqnarray}
a(p) &=& \frac{2^{(p-1)/2}\sqrt{3}\Gamma\left(\frac{3p-1}{12}\right)
     \Gamma\left(\frac{3p+19}{12}\right)\Gamma\left(\frac{p+5}{4}\right)}{8\pi^{1/2}(p+1)
     \Gamma\left(\frac{p+7}{4}\right)}, \\
F(p) &=&  \frac{2^{p+3}(p^2 + 4p + 11)}{(p+3)^2(p+5)(p+1)}
     \Gamma\left(\frac{p+5}{2}\right)\zeta\left(\frac{p+5}{2}\right).
\end{eqnarray}
Eqs.~\ref{eq4} and \ref{eq5} give the total emitted powers 
per volume per frequency in the rest frame. 
Given that both powers are diminished by the same dimming factor, 
the ratio of observed flux densities of the synchrotron and IC
emission, $S_{\rm Syn}/S_{\rm IC}$, is equal to $(dW_{\rm
  Syn}/d\nu_{\rm Syn} dt)/(dW_{\rm IC}/d\nu_{\rm IC} dt)$, and the 
strength of the magnetic field $B$ can be directly estimated.

We find $S_{\rm IC}<0.11~{\rm \mu Jy}$ from our limit on the
non-thermal hard X-ray emission for electrons with the IC emission
energy (frequency) of 12~keV ($\nu_{\rm IC}=2.9\times10^{18}(1+z)^{-1}$~Hz).  
As for $S_{\rm Syn}$, \citet{2007A&A...470L..25G} discovered extended
($\sim500$~kpc) radio emission in \object{RX~J1347.5--1145} based on
their low (18\arcsec) resolution VLA observations and higher
(2\arcsec) resolution data in the VLA archive.  They subtracted
contribution of discrete radio sources and estimated the total flux
density of the diffuse radio emission to be $S_{\rm Syn}=25$~mJy at
$\nu_{\rm Syn}=1.4(1+z)^{-1}$~GHz.  Combining these numbers with
Eqs.~(\ref{eq4}) and (\ref{eq5}), a lower bound on the magnetic field
strength in the ICM is obtained as $B > 0.007~{\rm \mu G}$ for
$p=2\Gamma-1=2$.  The estimation is sensitive to the assumption of
$p$: for example, in the case of $p=3$, $B>0.072~{\rm \mu G}$. 

This limit, though weak, is consistent with typical values found in
other clusters, $B\sim 0.1-1~{\rm \mu G}$, based on the {\it RXTE} and
{\it Beppo-SAX} observations \citep[e.g.,][]{2008SSRv..tmp...16R}.
Our result is also comparable to the recent {\it Suzaku} report on a
merging cluster \object{A3376} at $z=0.046$, $B>0.03~{\rm \mu G}$
\citep{2008arXiv0805.3582K}.  However, the previous measurements came
mostly from nearby ($z<0.1$) clusters as well as from some
medium-redshift clusters such as \object{A2163} at $z=0.20$
\citep{2006ApJ...649..673R} and the \object{Bullet cluster} at
$z=0.296$ \citep{2006ApJ...652..948P}.  Our work provides a constraint
on the cluster magnetic field strength at a higher redshift,
$z=0.451$.

\section{Summary}\label{sec:summary}
We have reported on the results from the analysis of the {\it Suzaku}
wide-band (0.5--10~keV with XIS and 12--60~keV with HXD/PIN) X-ray
spectroscopic observations of the most luminous X-ray cluster,
\object{RX~J1347.5--1145}, at $z=0.451$, in order to investigate the
temperature structure of the ICM, signatures of a recent violent
merger, as well as signatures of the non-thermal emission.

This cluster is known to contain a very hot gas clump that produces
the excess emission on top of the ambient gas that follows
more-or-less the conventional $\beta$ model
\citep{2001PASJ...53...57K,2002MNRAS.335..256A,2004PASJ...56...17K}.
The temperature of this gas clump has not been measured accurately
from the previous X-ray spectroscopic observations, as their
sensitivities degrade significantly beyond 10~keV.

The re-analysis of the {\it Chandra} data confirms the previous work,
and yields a poor limit on the temperature of the excess hot
component, $kT_{\rm ex}^{Chandra}=31.1^{+24.1}_{-12.6}$~keV (90\%
C.L.; statistical).  When the hard X-ray data from the {\it Suzaku}
XIS spectra in the 0.5--10~keV region and the HXD/PIN data in the
12--60~keV are combined with the {\it Chandra} data, we finally obtain
a good measurement of the temperature,
$25.3^{+6.1}_{-4.5}\,(^{+6.9}_{-9.5})$~keV (90\% C.L.; statistical and
systematic), which is in an excellent agreement with that derived from
the previous joint analysis of the SZ effect and X-ray imaging
observations \citep{2004PASJ...56...17K}. We stress that this is the
first time that the X-ray spectroscopic observations alone give a good
handle on such a high temperature gas component, which is made
possible by {\it Suzaku}'s unprecedented sensitivity over the wide X-ray
band.

We have found that the broad-band X-ray spectrum is explained better
by the thermal plasma model than by the non-thermal power-law model.
Thus, the present result confirms the presence of the hottest {\it
  thermal} gas in the cluster. The most likely explanation for this
phenomena is a recent violent merger with the collision velocity of
$\sim 4500$~km/s, similar to the one found in the \object{Bullet
  cluster}\citep{2007ApJ...661L.131M,2007MNRAS.380..911S,
2007arXiv0711.0967M,2008MNRAS.384..343N}.

The upper bound on the non-thermal flux in the 12--60~keV band,
$F_{\rm HXR}<8\times10^{-12}~{\rm erg\,s^{-1}\,cm^{-2}}$, yields, when
combined with a recent discovery of the radio mini halo in this
cluster by \citet{2007A&A...470L..25G}, a lower bound on the ICM
  magnetic field strength, $B>0.007~{\rm \mu G}$ for the electron
  index of 2.  These constraints provide valuable information on the
non-thermal nature of the cluster gas and the particle acceleration in
a high-redshift universe.  However, the accuracy of the present
  measurement of the hard X-ray emission is limited by the sensitivity
  of the instrument.  Suppose that the magnetic field in
  \object{RX~J1347.5--1145} is comparable to that of the other
  clusters, $B\sim 0.1~{\rm \mu G}$, a 50-fold higher  
sensitivity is required to detect non-thermal hard X-ray emission from this cluster.
We expect that the future X-ray missions \citep[e.g., {\it NeXT};][]{2008arXiv0807.2007T} 
will determine the non-thermal X-ray flux accurately and draw a more complete picture. 

Before the advent of the {\it Suzaku} satellite, it was difficult to
measure the high-temperature ($\gg 10$~keV) gas in the X-ray band
reliably, due to the limited sensitivity.  The present paper
demonstrates the power of the broad-band X-ray spectroscopy in the
study of gas heating due to the cluster merger, which is nicely
complementary to the observations of the SZ effect. In the near future
we expect the number of samples of merging clusters observed with {\it
  Suzaku} to increase rapidly, which would shed new light on the
physics of violent mergers.
 
\begin{acknowledgements}
  We are grateful to the {\it Suzaku} team members for the operation
  and the instrument calibrations. We also thank H. B\"{o}hringer,
  T. Ohashi, K. Masai, N.Y. Yamasaki, K. Mitsuda for discussions.  NO
  acknowledges support from the Alexander von Humboldt Foundation in
  Germany. This work is supported in part by the Grant-in-Aid by the
  Ministry of Education, Culture, Sports, Science and Technology,
  19740112 (NO) and 18740112 (TK).  EK acknowledges support from an
  Alfred P. Sloan Research Fellowship.
\end{acknowledgements}

\bibliographystyle{aa}
\bibliography{rxj}

\appendix
\section{The {\it Suzaku}-{\it Chandra} cross calibration}\label{appendix:suzaku_chandra}
Because the temperature measurement of very hot gas based on the
spectral fit is sensitive to a subtle change of the spectral slope, a
precise calibration of the effective area is critical. 
In the analysis of XIS+HXD spectra
(\S\ref{sec:xis+hxd_analysis}), we have modeled the cluster average
emission with {\it Chandra}. Thus the cross-calibration between {\it
  Suzaku}/XIS and {\it Chandra}/ACIS-I is mandatory.

If the global cluster spectrum is accumulated from the $r<5'$ circular
region with the {\it Chandra}/ACIS-I data, the APEC model fitting gave
$kT=14.4(13.6-15.3)$~keV, $Z=0.45(0.38-0.52)$,
$K=1.47(1.45-1.49)\times10^{-2}$ ($\chi^2/{\rm dof}=552.8/426$), where
$z=0.451$ and $N_{\rm H}=4.85\times10^{20}~{\rm cm^{-2}}$ are
adopted. In comparison with the best-fit XIS parameters
(Table~\ref{tab2}), the temperature and the normalization are higher
by about 1.5~keV and 7\% for {\it Chandra}. We discuss below the possible
reasons in the light of the {\it Suzaku} and {\it Chandra} calibration
uncertainties.

\begin{enumerate}
\item {\it Suzaku} XIS\\ Since the spatial extent of X-ray emission from
\object{RX~J1347.5--1145} is small compared to the PSF of
{\it Suzaku}, it can be treated as a point source (we have confirmed that
the spectral fit with an auxiliary file generated for a point source
at the HXD-nominal position yields statistically consistent spectral
parameters with those listed in Table~\ref{tab2}).  According to the
latest calibration of XIS and PIN, the best-fit power-law parameters
describing the Crab spectra are the photon index
$\Gamma=2.073\pm0.006$ and the normalization $9.21\pm 0.1~{\rm
photons\,cm^{-2}s^{-1}keV^{-1}}$, which are by about $1.5$\% and 5\%
lower than the `standard' values \citep[$\Gamma=2.10$, the
normalization $=9.7$;][]{1974AJ.....79..995T}.

To simply simulate the effect of $\Gamma$, we multiply the APEC model
by the PLABS model, $E^{-\alpha}$, with $\alpha=-0.03$ and fit it to the
XIS+PIN spectra. The best-fit parameters are obtained to be
$kT=11.85$~keV, $Z=0.30$~solar, and $K=1.35\times10^{-2}$
($\chi^2/{\rm dof}=1343/1219$). This indicates that the {\it Suzaku}
temperature measurement has a systematic error at the level of
$\sim1$~keV, however, the discrepancy between {\it Suzaku} and {\it Chandra}
becomes even larger in this case. Thus the uncertainty of the {\it Suzaku}
effective area is unlikely to be the major source of the discrepancy.

\item {\it Chandra} ACIS-I \\ Based on a comparison of cluster
  temperatures derived from {\it Chandra} and {\it XMM-Newton},
  \cite{2008A&A...478..615S} reported that the {\it Chandra}
  temperature tends to be by 1--2~keV higher than that of {\it
    XMM-Newton} particularly for the clusters with
  $kT\gtrsim5$~keV. Furthermore, the proceedings of the {\it Chandra}
  Calibration Workshop by David et al.
  \footnote{http://cxc.harvard.edu/ccw/proceedings/07\_proc/presentations/david/}
  indicated the same tendency as well as that the temperature obtained
  from the fit to the 2--6~keV continuum spectra is systematically
  higher by about 2~keV than that inferred from the iron line ratios for
  hot clusters.  Thus we consider that the above results suggest that
  the {\it Chandra} effective area has a systematic error of the level
  of $\Delta kT = +1\sim+2$~keV for hot ($kT\gtrsim5$~keV) clusters.

  For example, fitting the {\it Chandra} data with the APEC model
  multiplied by the $\alpha=-0.03$ PLABS model leads to
  $kT=12.89$~keV, $Z=0.40$~solar, and $K=1.44\times10^{-2}$
  ($\chi^2/{\rm dof}=569/426$). Although the metal abundance is
  marginally higher, the temperature agrees with that of XIS.

  Then by simultaneously fitting the {\it Chandra} and XIS spectra
  under the APEC$\times$PLABS model and determined $\alpha$ and the
  relative normalization factor to be $-0.017(-0.031\sim -0.003)$ and
  $1.07(1.06-1.08)$, respectively. The best-fit APEC parameters are
  $kT=12.92$~keV, $Z=0.35$~solar, and $K=1.37\times10^{-2}$
  ($\chi^2/{\rm dof}=1890/1636$). The result is plotted in
  Fig.~\ref{figa1}

%%%%%%%%%%%%%%%%%%%%%%%%%%%%%%%%%%%%%%%%%%%%
\begin{figure}
\centering
\rotatebox{270}{\scalebox{0.33}{\includegraphics{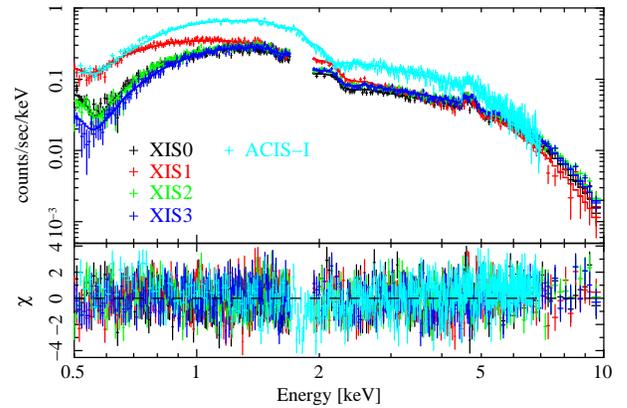}}}
\caption{{\it Suzaku}/XIS+{\it Chandra}/ACIS-I spectra of
  \object{RX~J1347.5--1145} ($r<5\arcmin$) simultaneously fitted with
  the APEC$\times$PLABS model}
\label{figa1}
\end{figure}
%%%%%%%%%%%%%%%%%%%%%%%%%%%%%%%%%%%%%%%%%%%%

\end{enumerate}

Based on the above discussion, in order to take into account the
difference of the calibration between the two satellites, we assume in
\S\ref{sec:xis+hxd_analysis} the {\it Chandra} normalization factor
relative to XIS of $1/1.07=0.93$ and multiply the spectral model
derived with {\it Chandra} by the PLABS model with
$\alpha=0.02(0-0.03)$.

%\listofobjects

\end{document}